# Preventing quartz–NaI adhesion in Bridgman growth using ammonium iodide

*Lam Tan Truc[1], N. T. Luan[2], Gul Rooh[1,3], O. Gileva[4], K.A. Shin[4], H.S. Lee[4], A. Iltis[5], C.R. Byeon[4], C.H. Lee[4], H.J. Kim[1,2*]*

[1]Department of Physics, Kyungpook National University, Daegu, 41566, Republic of Korea

[2]Center for High Energy Physics, Kyungpook National University, Daegu, 41566, Republic of Korea

[3]Department of Physics, Abdul Wali Khan University Mardan, 23200, Pakistan

[4]Center for Underground Physics, Institute for Basic Science (IBS), Daejeon 34126, Republic of Korea

[5]Damavan Imaging, Troyes, 10430, France

*Email: hongjoo@knu.ac.kr (Corresponding author)

## Abstract:

Adhesion between NaI(Tl) single crystals and the walls of the quartz ampoules remains a major limitation for Bridgman growth under sealed conditions, particularly for applications requiring ultra-radiopure scintillators, where sealed handling is required. In this study, ammonium iodide ($NH_4I$) was used as an additive with the aim of generating HI in situ, which suppresses the NaOH–$SiO_2$ reaction that forms adhesive sodium–silicate phases. Small-sized crystals (Φ8 mm) were first grown to determine the $NH_4I$ concentration required to eliminate adhesion. The optimized condition was used to grow a large-sized NaI(Tl) crystal. A crack and bubble-free NaI(Tl) crystal with dimensions of Φ3-inch × 3-inch was successfully grown. The crystal exhibited 59,000 ph/MeV light output, which is higher than other commercial NaI(Tl) crystals used as references in this study.

## I. Introduction

NaI(Tl) is widely used as a scintillator material in radiation detection [1-2], and the demand for large, high-purity crystals continues to increase as detector systems become larger and more sensitive [3]. For dark matter searches and low-background experiments, the radiopurity of both the starting materials (NaI powder, TlI dopant) and the materials comprising the growth environment (e.g., quartz ampoules and residual contaminants within the sealed system) is particularly important [4–7,22], as both can contribute to intrinsic and surface-related radioactive backgrounds in the final crystal. Sealed growth configurations are often preferred to limit exposure to external contaminants. In addition, achieving a higher light output from the grown NaI(Tl) crystals is crucial for enhancing dark-matter detection sensitivity.

Several industrial techniques, such as the Czochralski and Kyropoulos methods, can produce large NaI(Tl) crystals and are widely used in commercial manufacturing [23]. These methods typically employ open crucible systems (e.g., platinum crucibles) rather than sealed quartz ampoules, thereby avoiding direct interaction between the melt and $SiO_2$ and eliminating the associated adhesion problem. In contrast, the Bridgman method remains attractive for laboratory-scale growth due to its simplicity, low equipment cost, and the ready availability of quartz ampoules compared with a platinum crucible, as well as the ability to maintain an isolated environment suitable for handling hygroscopic and ultra-radiopure materials.

Although attractive, there is a persistent limitation of Bridgman growth in quartz ampoules - the strong adhesion of NaI(Tl) to the inner wall of the quartz ampoule, which leads to cracking of the ingot and material loss during extraction [2,6,8-9]. Furthermore, ampoule cracking may expose the hygroscopic NaI(Tl) surfaces and complicate post-growth handling. The adhesion is attributed to reactions between residual NaOH and the quartz surface. NaOH is commonly associated with NaI powders due to their hygroscopic nature, where exposure to moisture can lead to partial hydrolysis during production, storage, or handling. The exact NaOH content depends on the powder history and moisture exposure prior to use. In this work, the starting NaI powder [24] was handled and dried under vacuum prior to loading; however, such treatment removes moisture but may not fully eliminate pre-existing NaOH. Because high-purity NaI powder is costly and often limited, maximizing material yield during crystal growth is an important practical consideration. In Bridgman growth using sealed quartz ampoules, adhesion/cracking/air exposure can result in partial loss of the grown boule, with difficulty in melt recovery. This contrasts with other growth techniques, such as the Kyropoulos method, where residual melt can often be reused. Therefore, suppressing adhesion is not only beneficial for crystal quality but also critical for improving material utilization efficiency in Bridgman-grown NaI(Tl) crystals.

Carbon-coated quartz ampoules [6,8,12] and graphite-based crucibles [27] have been widely used to avoid adhesion between molten NaI and quartz surfaces in Bridgman growth. However, previous studies, particularly B. Suerfu et al. [6], observed residual carbon flakes embedded within NaI(Tl) crystals and suggested that such inclusions may contribute to optical absorption and reduced light yield. Carbon-based materials such as graphite exhibit strong optical absorption in the visible spectral range, with absorption coefficients on the order of several $\mu m^{-1}$ around 400–500 nm [28], corresponding to sub-micrometer attenuation lengths. Since the scintillation emission of NaI(Tl) lies in the 410–430 nm range, even thin carbon inclusions or surface residues may lead to appreciable loss of scintillation photons. This provides a plausible mechanism by which carbon-related features may degrade light collection efficiency. In this work, crystals grown using carbon-coated ampoules exhibited approximately 70% of the light yield compared to those grown without carbon under identical source powders and growth conditions. This controlled comparison suggests that carbon coating can negatively affect scintillation performance in our growth configuration. In addition, carbon coating introduces extra preparation steps and may lead to variability in coating quality, as well as unintended incorporation of contaminated materials into the growth environment [13]. For applications requiring ultra-low background levels, introducing non-native materials into the melt is typically avoided.

The ammonium iodide ($NH_4I$)-based additive approach, which modifies the chemical environment inside the sealed ampoule, offers an alternative method to prevent adhesion without introducing a significant mass of foreign material that could potentially contain impurities. Upon heating, $NH_4I$ is expected to decompose into $NH_3$ and HI gases, and this released HI may react with residual NaOH present in the system, converting it back to NaI. This mechanism would suppress the formation of sodium–silicate phases at the quartz interface, which are responsible for strong adhesion between the melt and the ampoule wall. Because the additive is required only in very small amounts and can be incorporated directly into the starting charge, this approach eliminates the need for additional equipment or separate preparation steps.

In this work, we examine the effect of $NH_4I$ on adhesion behavior during Bridgman growth of NaI(Tl) in quartz ampoules. To our knowledge, the use of $NH_4I$ as an in-situ chemical additive for suppressing NaI–quartz adhesion has not been previously reported. Small-diameter crystals were first grown to determine the optimal $NH_4I$ concentration required to eliminate sticking, and the optimized condition was subsequently applied to the growth of larger size NaI(Tl) crystals. The influence of the additive on bubble formation and the corresponding adjustments to the growth rate were also investigated.

## Experiment

### 1.1. Crystal growth

To ensure identical growth conditions, all samples in this study, including both small- and large-size crystals, were grown by the vertical Bridgman technique in quartz ampoules using the same custom-built furnace. The furnace design shown in Fig. 1 was based on the configuration reported by Phan et al. (2022) [14].

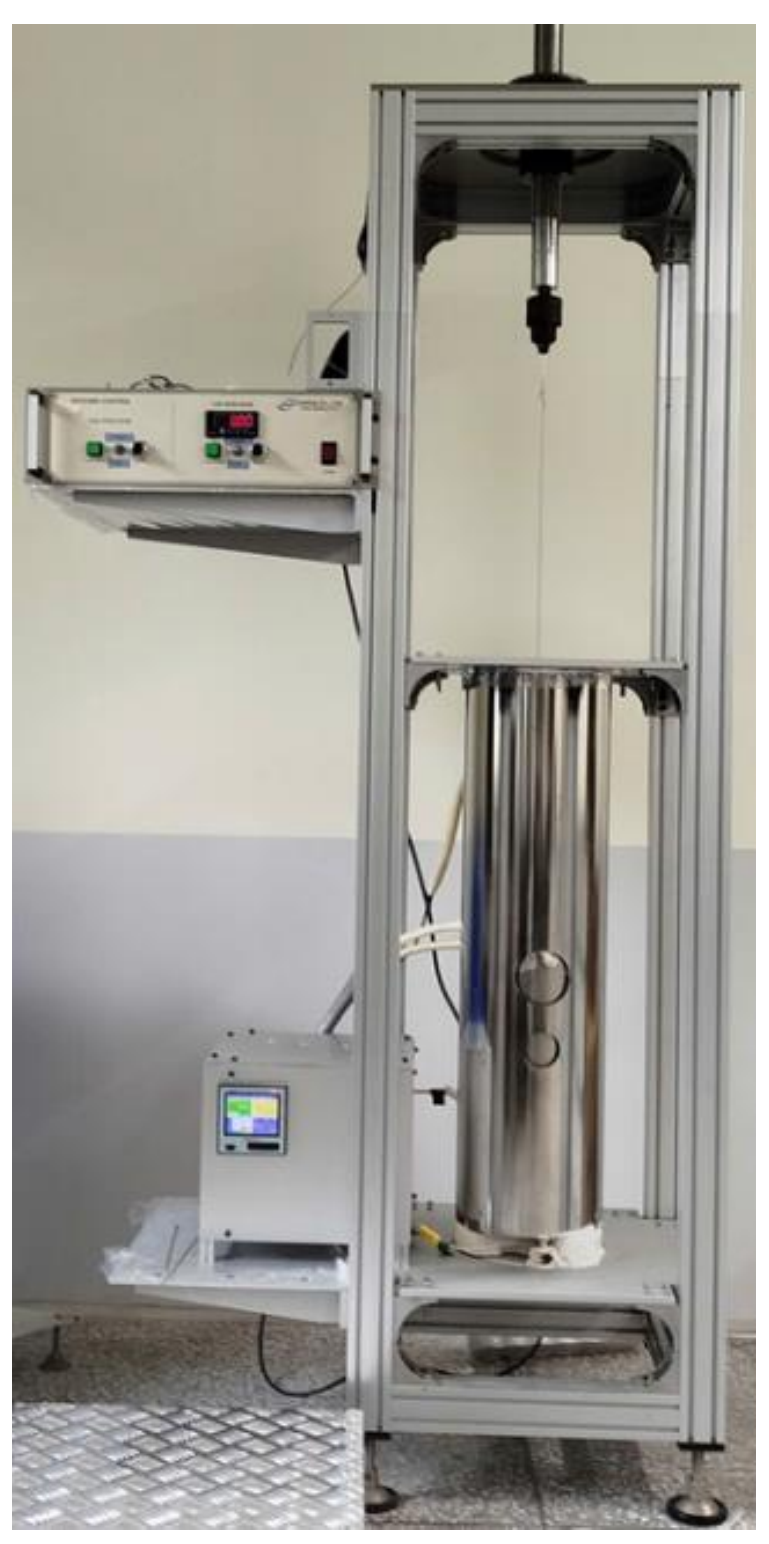

***Fig. 1.*** *Vertical Bridgman furnace used to grow crystals.*

Five quartz ampoules (outer diameter: 1 cm) were prepared with $NH_4I$ concentrations of 0, 75, 100, 250, and 500 mg/kg (ppm) of NaI powder to study adhesion behavior. An additional carbon-coated ampoule was prepared for comparison of scintillation performance (Fig. 2(a)). High-purity $NH_4I$ (5N, Sigma-Aldrich) and ultra-radiopure NaI powder supplied by the COSINE-100 collaboration [7] were used, with 0.1 mol% TlI as the activator. All powder handling and loading processes were carried out in a nitrogen-purged glovebox with the moisture level ($H_2O$) maintained below 300–400 ppm at 25 °C to minimize exposure to humidity. Prior to loading, the quartz ampoules were cleaned by sequential rinsing with 99.9% ethanol and deionized water (three times each) to remove surface contaminants. This was followed by chemical etching using a 0.5 M HF solution at 60 °C for 1 hour. After etching, the ampoules were thoroughly rinsed with deionized water and dried under controlled conditions for 2–3 days. The cleaned ampoules were then transferred into the glovebox, where $NH_4I$ powder was first loaded, followed by NaI and TlI powders. The loaded ampoules were then coupled to a high-vacuum (below $10^{-6}$ Torr), dried at 140 °C for 24 hours, and later sealed using a propane torch. The overall preparation procedure was adapted from Ref. [6], with modifications to suit the present work.

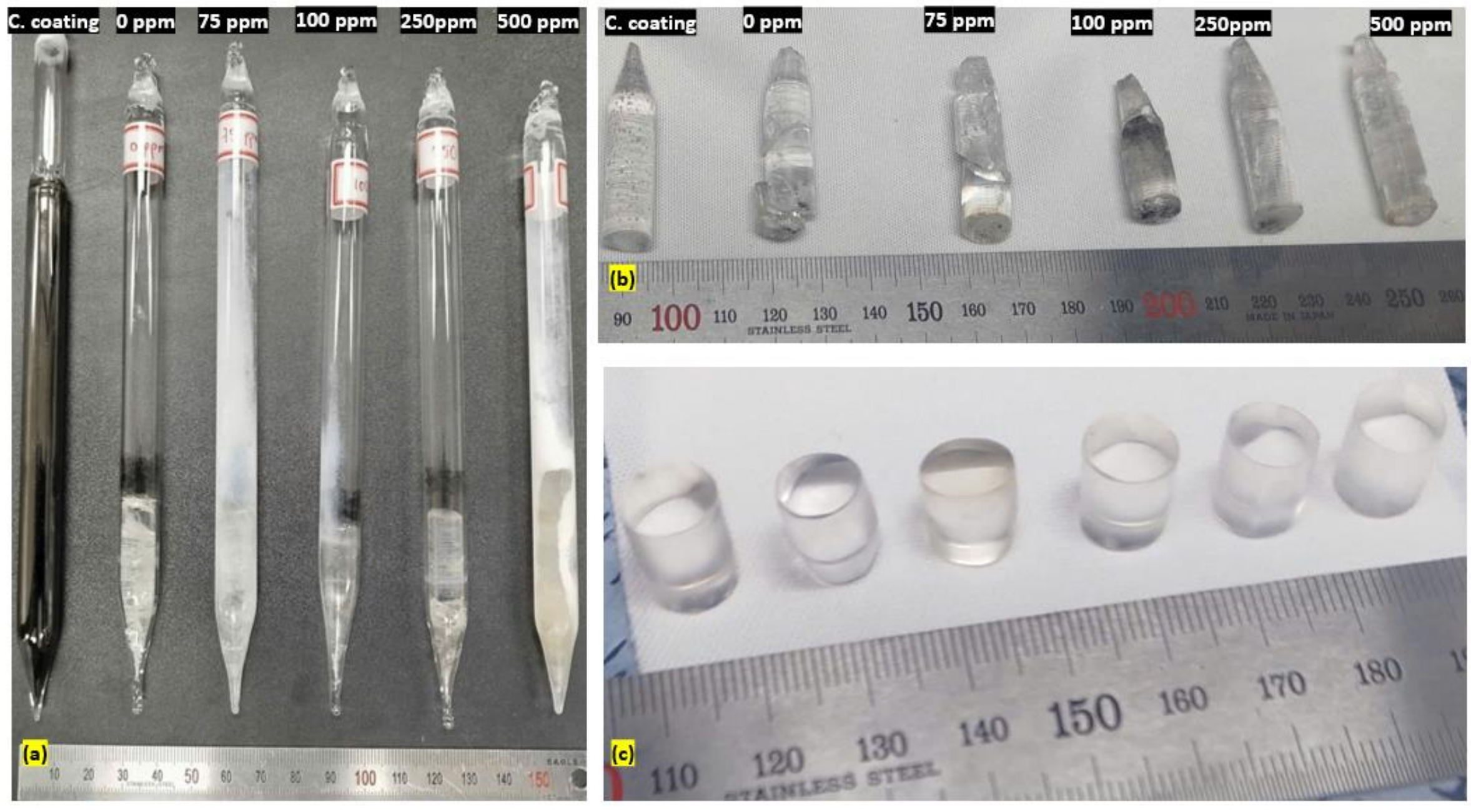

***Fig. 2. (a)*** *Small-sized ampoules after growth;* (b) *crystals removed from ampoules: one carbon-coated, one without* $NH_4I$*, and four with different* $NH_4I$ *concentrations;* ***(c)*** *crystals after polishing.*

All six small ampoules were grown simultaneously in the same furnace. At the start of growth, both the hot and cold zones were set to 700 °C to fully melt the NaI mixture and promote convection for uniform thallium distribution. After 12 h, the hot zone was maintained at 50 °C above the NaI melting point, while the cold zone was adjusted to maintain a 10 °C/cm temperature gradient. After solidification, the furnace was cooled to room temperature at a rate of 8 °C/h.

After growth, all ampoules were transferred to a glovebox. Inside the glovebox, the ampoules were cut with a diamond-coated wire saw to preserve the original inner and crystal surfaces for Scanning Electron Microscopy (SEM), Energy-dispersive X-ray spectroscopy (EDS), and X-ray Diffraction (XRD) analysis. The ampoule's inner surfaces were used for SEM/EDS analysis, while the crystal tails (Φ8 mm × 10 mm) were polished for scintillation measurements. A small portion of each crystal was ground into powder inside the glovebox and then sealed for XRD analysis. The cylindrical tails were polished using fine sandpaper without mineral oil. The polished crystals prepared for characterization are shown in Fig. 2 (c).

For SEM/EDS analysis, the NaI(Tl) samples were gold-coated using a sputter coater in a low-humidity environment to prevent moisture absorption. SEM measurements were then carried out under vacuum.

### 1.2 Characterization

The XRD measurements and scintillation characterization procedures followed the same configuration as reported in [15], with minor modifications. In this work, a Hamamatsu R6233-100 photomultiplier tube was operated at −800 $V_{bias}$ to record the scintillation decay profiles and to acquire pulse-height spectra for light-yield and energy-resolution evaluation. The decay curves were fitted using a two-component exponential function, while the energy resolution was determined from the full width at half maximum (FWHM) obtained by Gaussian fitting of the 662 keV photopeak. For light yield calculation, the reference crystals included $8 \times 8 \times 8$ $mm^3$ ASCII-2 NaI(Tl) crystals for small-size samples [16], and commercial NaI(Tl) detectors from EPIC (Φ2-inch × 2-inch) and Alpha Spectra (Φ3-inch × 3-inch) for large-size samples (Fig. 14(c) and (d)).

SEM was performed using an FEI Versa 3D dual-beam system to characterize the surface morphology of the sample. Images were acquired in secondary-electron (SE) mode using the Everhart-Thornley detector (ETD) at an accelerating voltage of 20 kV. A working distance of approximately 10–15 mm and a spot size of 5.0 were used to balance image resolution and beam stability. The sample was mounted without stage tilt (0°), and micrographs were recorded at magnifications ranging from 50× to 1000×, with a field of view of 1 mm to 50 µm. These conditions are typical for the surface characterization of insulating/partially coated inorganic materials, such as our NaI(Tl) and quartz samples.

EDS analysis was performed using the same FEI Versa 3D dual-beam system equipped with an energy-dispersive X-ray spectroscopy detector. Measurements were conducted under an accelerating voltage of 20 kV, consistent with the SEM imaging conditions. The analysis focused on regions of interest at the inner ampoule surface where adhesion or interfacial reactions were observed. EDS provides elemental composition only; therefore, Na-containing species (e.g., NaI, NaOH, or sodium–silicate phases) cannot be distinguished unambiguously. The results are thus interpreted qualitatively.

To further investigate the interfacial reaction mechanism between NaOH and quartz, additional ampoule tests were performed at 200 °C and 700 °C using only NaOH and NaOH +

$NH_4I$ (Fig. 3). The resulting samples were analyzed using SEM/EDS under the same conditions described above.

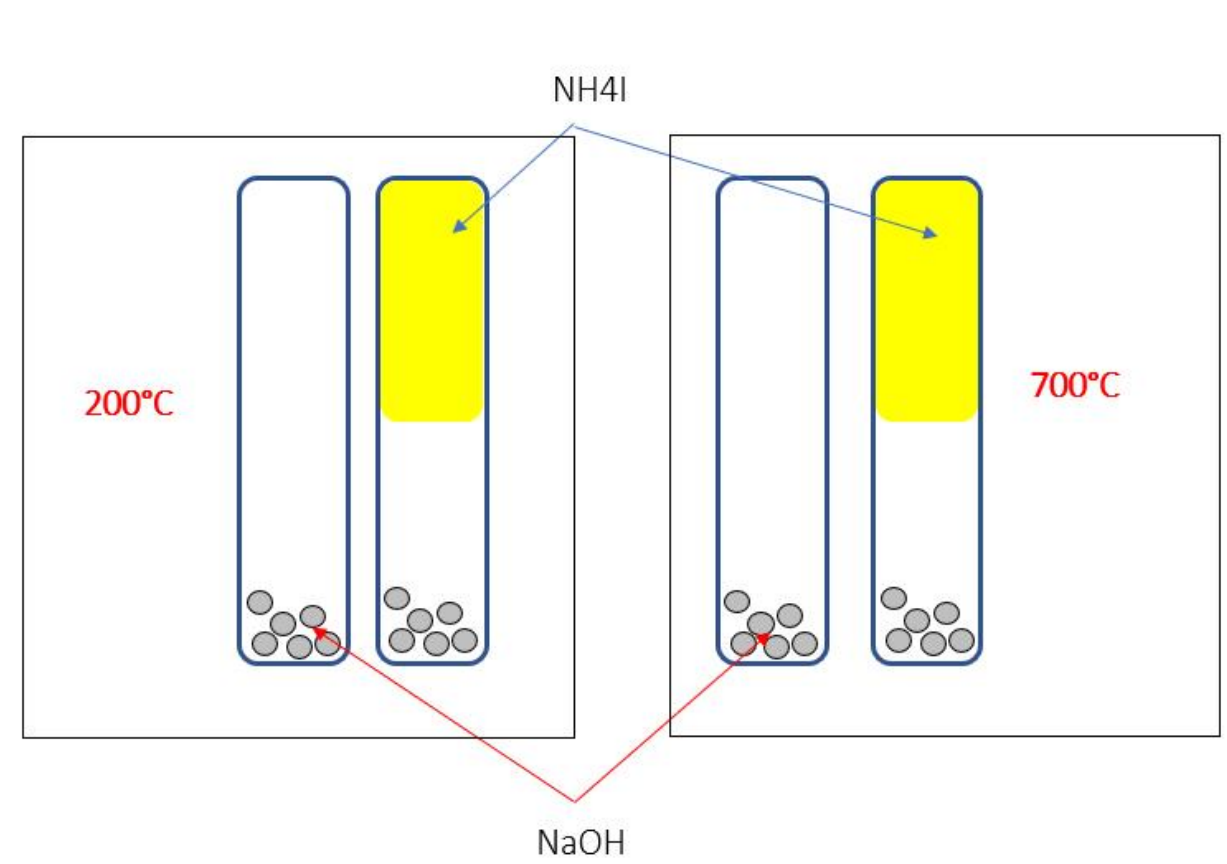

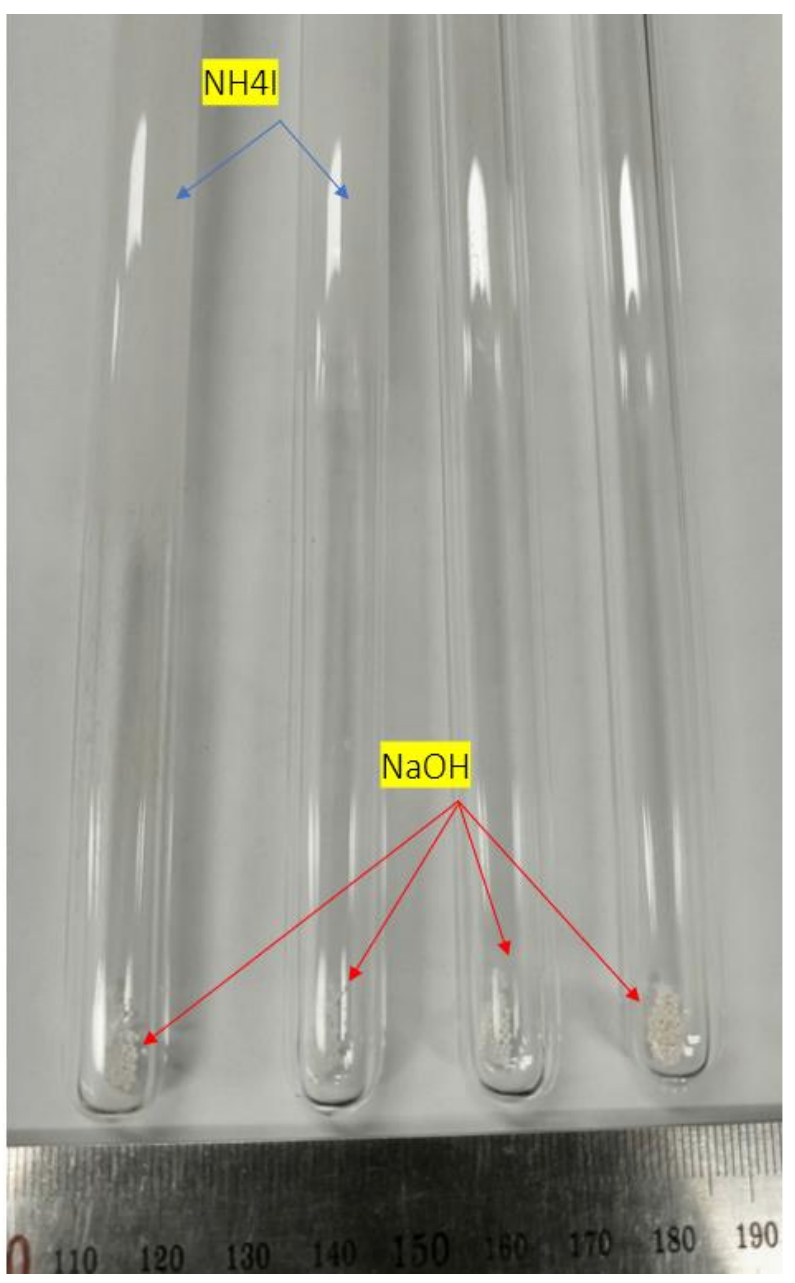

***Fig. 3.*** *Interfacial reaction between NaOH and quartz with and without $NH_4I$ at 200 °C and 700 °C*

# II. Results and discussion

## 2.1. XRD analysis

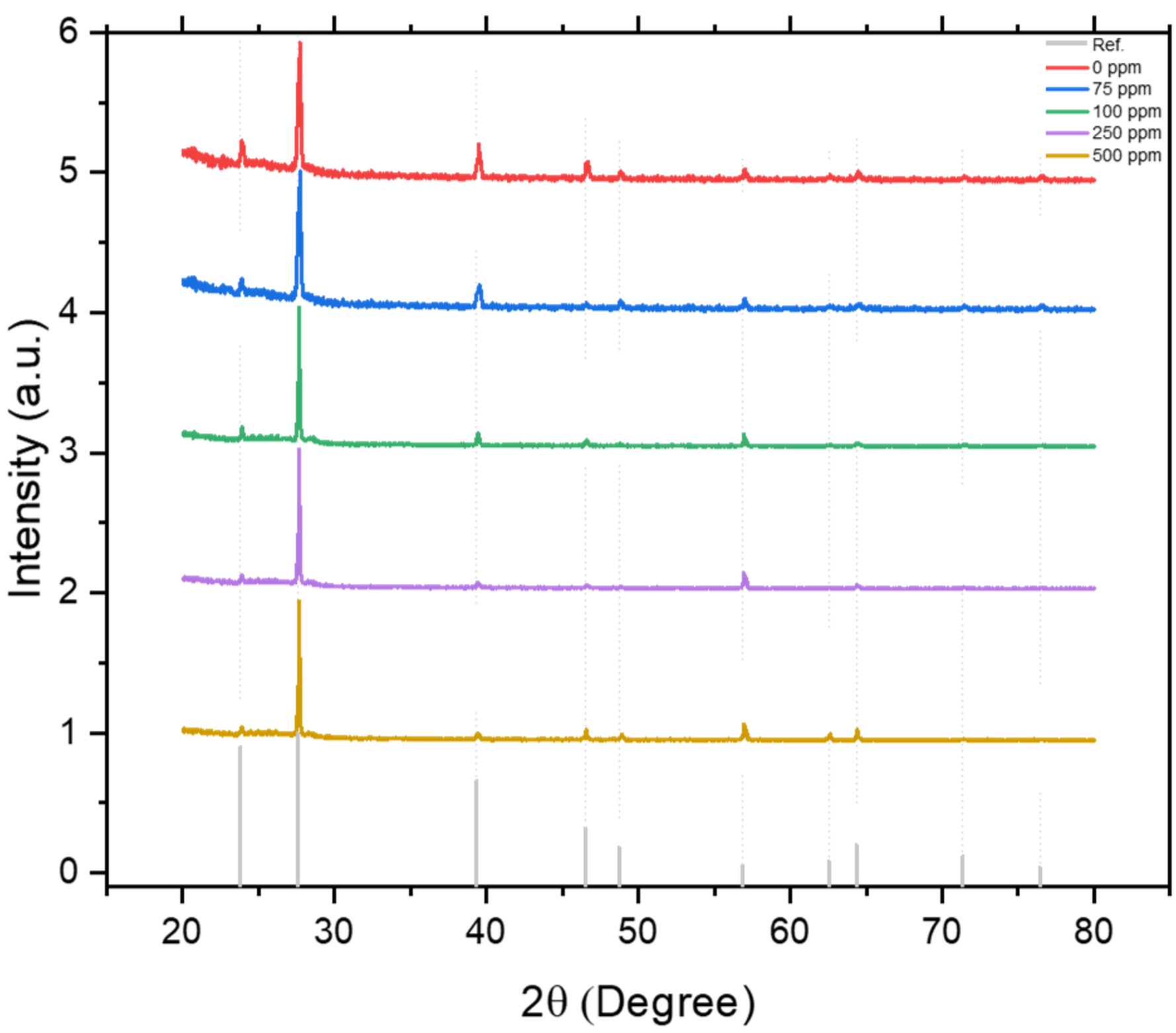

***Fig. 4.*** *XRD pattern of small-sized crystals compared with the NaI reference pattern.*

As shown in Fig. 4, all samples display identical diffraction peaks corresponding to cubic NaI, matching the ICDD reference pattern 00-001-0715. No additional peaks were observed, confirming that the addition of $NH_4I$ did not change the single-phase NaI structure.

### 2.2. X-ray induced luminescence

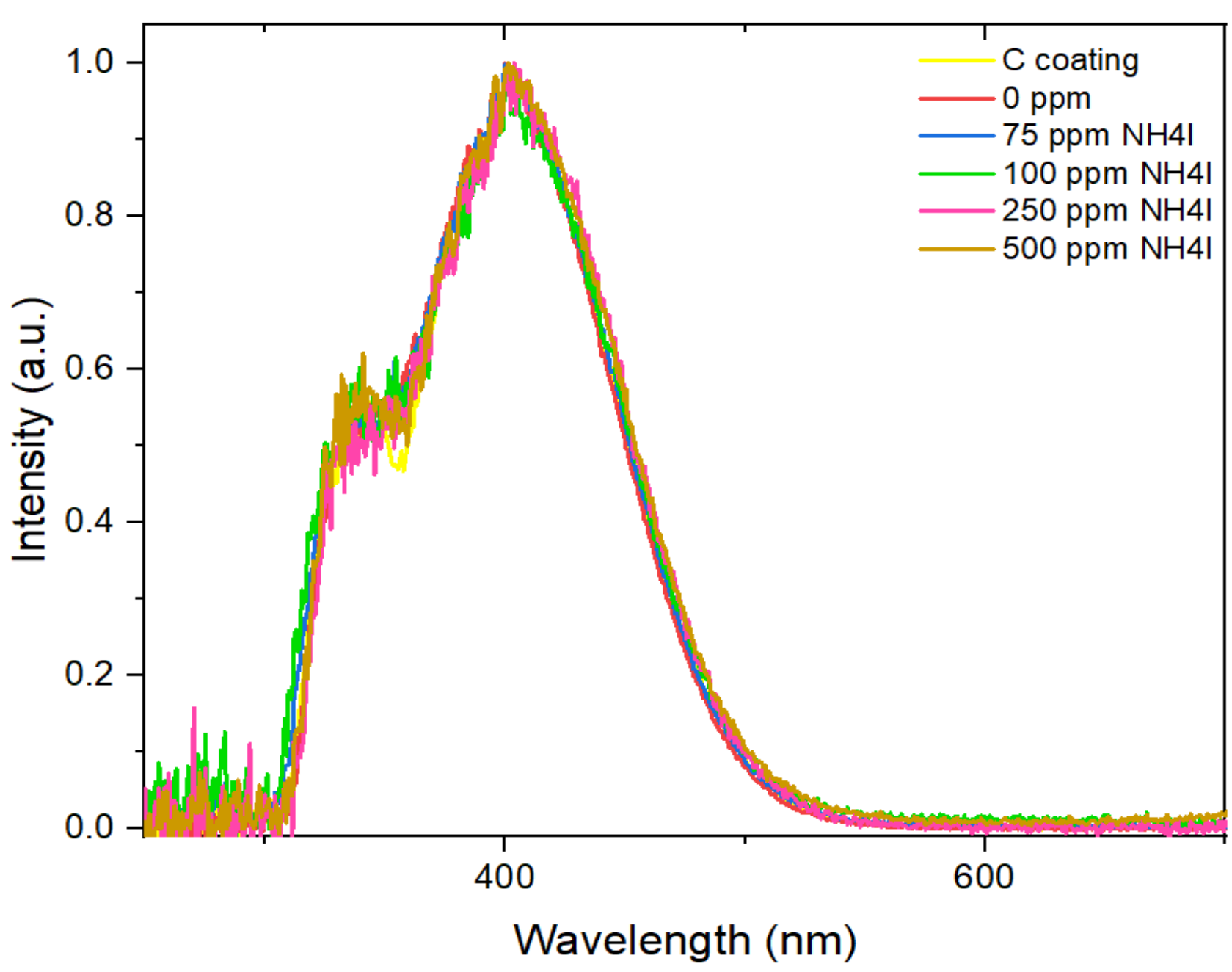

***Fig. 5.*** *X-ray luminescence spectrum of small-sized crystals at room temperature.*

The X-ray–excited emission spectra of the NaI(Tl) crystals are shown in Fig. 5. All samples exhibit a main emission peak at 430 nm and a weaker band near 380 nm, consistent with literature values [16,19]. These results confirm that the addition of $NH_4I$ does not affect the luminescence center. The identical spectra also suggest no strong variation of thallium distribution within the crystal lattice, since the emission wavelength is known to vary with thallium concentration, as reported by Eby and Jentschke [20].

### 2.3. Pulse height spectra/Decay

Table 1 summarizes the scintillation properties of all small-sized NaI(Tl) samples under 662 keV γ-ray excitation from a $^{137}Cs$ source, compared with the 8 × 8 × 8 $mm^3$ reference crystal. The 250-ppm $NH_4I$ sample showed the best overall performance, while the carbon-coated sample had the lowest light yield and worst energy resolution. Among the $NH_4I$-treated samples, the 75-ppm sample showed the lowest light output, and the 500-ppm sample exhibited a poor energy resolution. Considering the ±10 % systematic uncertainty, no clear trend in either light yield or energy resolution was observed, indicating that $NH_4I$ did not affect the intrinsic scintillation performance of NaI(Tl). Instead, $NH_4I$ primarily influences adhesion: at higher concentrations (250–500 ppm), sticking was completely eliminated, allowing the crystals to be removed from the ampoules intact.

***Table 1**: The scintillation properties at 662 keV of small-sized NaI(Tl) samples and reference 8 x 8 x 8 $mm^3$ crystal.*

| Sample | Light yield (ph/MeV) | Energy Resolution (%) | Decay Time (ns) | Adhesion |
|---|---|---|---|---|
| 8 x 8 x 8 $mm^3$ | 52,000 ± 5,200 [16] | 6.8 ± 0.1 | 226 ns (88 %) | - |
| Carbon coating | 36,300 ± 3,630 | 11.0 ± 0.2 | 231 ns (89 %) | no |
| 0 ppm | 46,400 ± 4,640 | 8.1 ± 0.1 | 214 ns (86 %) | strong |
| 75 ppm | 39,400 ± 3,940 | 10.2 ± 0.2 | 213 ns (89 %) | medium |
| 100 ppm | 42,200 ± 4,220 | 10.0 ± 0.2 | 223 ns (89 %) | little |
| 250 ppm | 51,300 ± 5,130 | 7.8 ± 0.1 | 213 ns (88 %) | no |
| 500 ppm | 43,700 ± 4,370 | 10.7 ± 0.2 | 220 ns (89 %) | no |

For scintillation characterization, samples were taken from the cylindrical tail region of each boule whenever possible. However, as shown in Fig. 2 (b), several crystals - particularly those with 0, 75, and 100 ppm $NH_4I$ - were partially broken, so samples from these groups were cut slightly closer to the middle section of the boules. During Bridgman growth, because thallium has an effective segregation coefficient $k_{eff} < 1$, it preferentially remains in the melt and accumulates in the last-to-freeze (tail) region during growth [21]. Nevertheless, the measured fast decay components for all samples remained consistent within the ±10 % systematic uncertainty (Fig. 8 (a)), confirming the acceptable uniform thallium distribution along the length of the boules.

Although all small-sized samples had similar dimensions and were taken from comparable positions in the boules, some surfaces - particularly in the carbon-coated sample - showed dents and scratches. Additional polishing was required to remove surface defects, which occasionally resulted in slight variations in the final cylindrical shape. Differences in the thickness of the Teflon wrapping and the optical grease layer between the crystal and the PMT may also contribute to variations in light yield and energy resolution.

Fig. 6 compares the pulse height spectra of the carbon-coated sample, the 250 ppm $NH_4I$ sample, and the 75 ppm $NH_4I$ sample with the 8 × 8 × 8 mm³ reference crystal under 662 keV γ-ray excitation. The 250-ppm sample exhibited a Gaussian-fitted photopeak with improved resolution, confirming stable light collection and uniform Tl distribution, shown in Fig. 7.

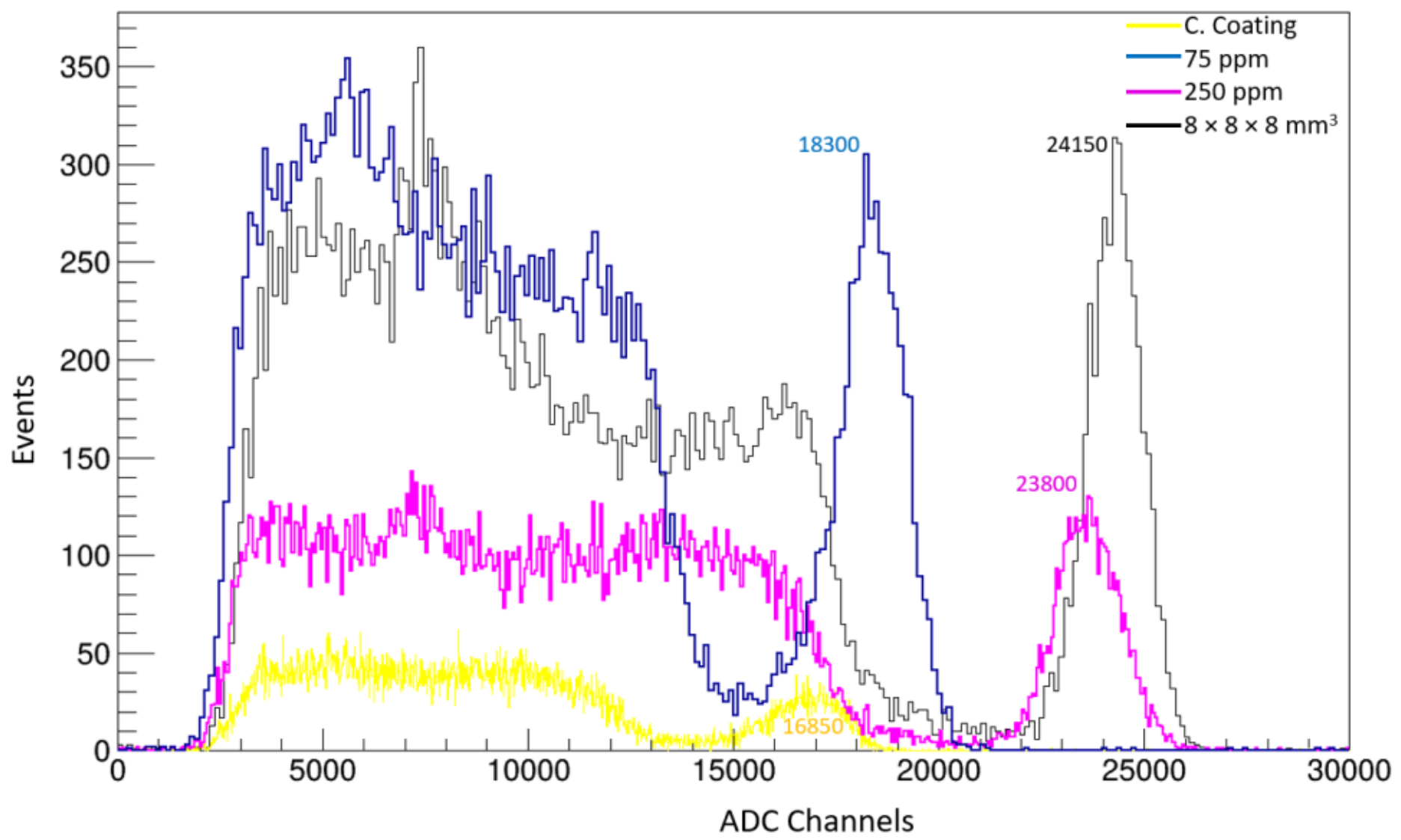

***Fig. 6.*** *Pulse height spectra at 662 keV under ¹³⁷Cs γ-ray excitation for the carbon-coated, 75 ppm, and 250 ppm $NH_4I$ samples compared with the 8 × 8 × 8 mm³ reference crystal.*

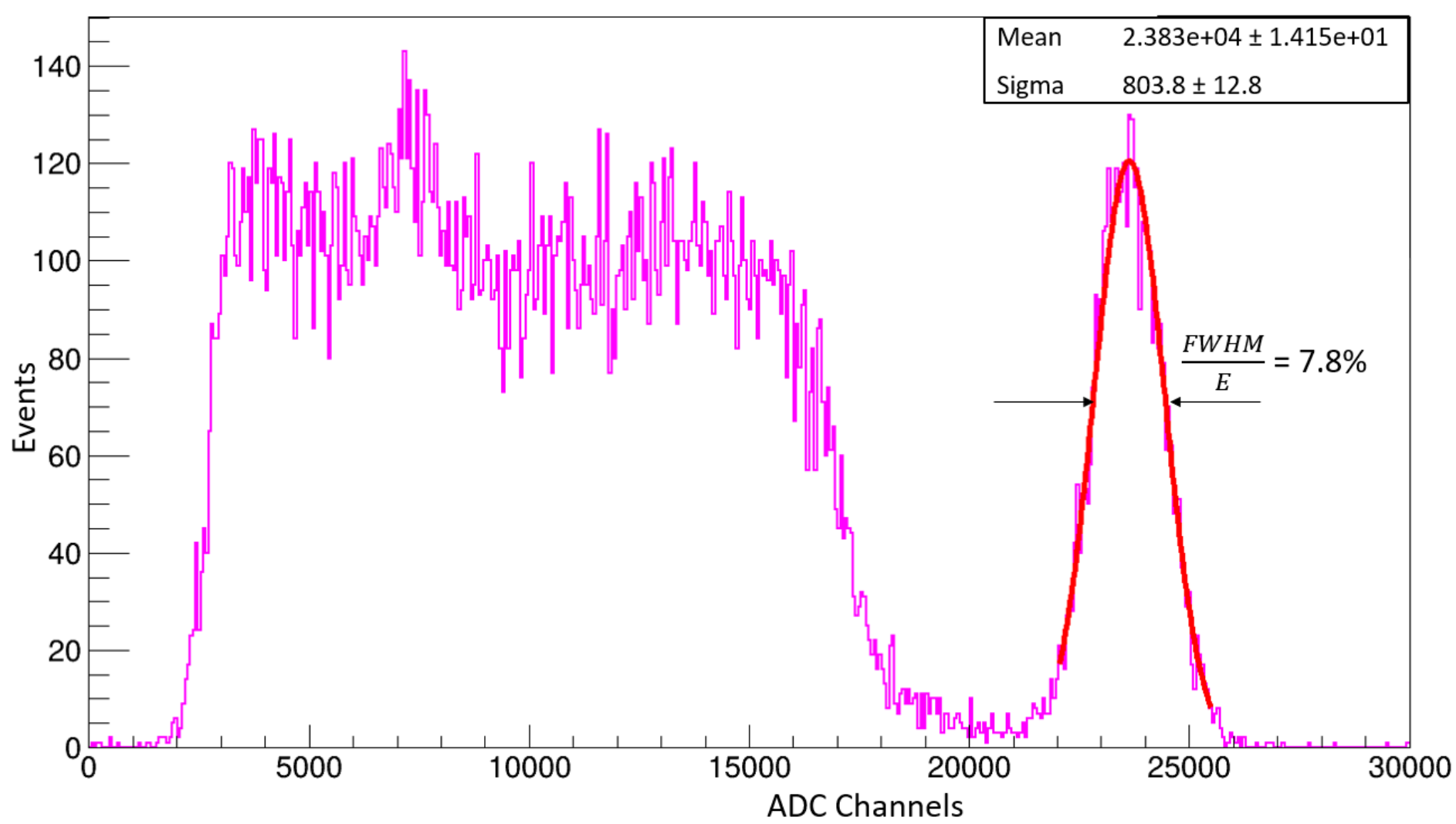

***Fig. 7.*** *Pulse height spectrum and energy resolution (FWHM/E) of the 250 ppm $NH_4I$ sample under ¹³⁷Cs excitation.*

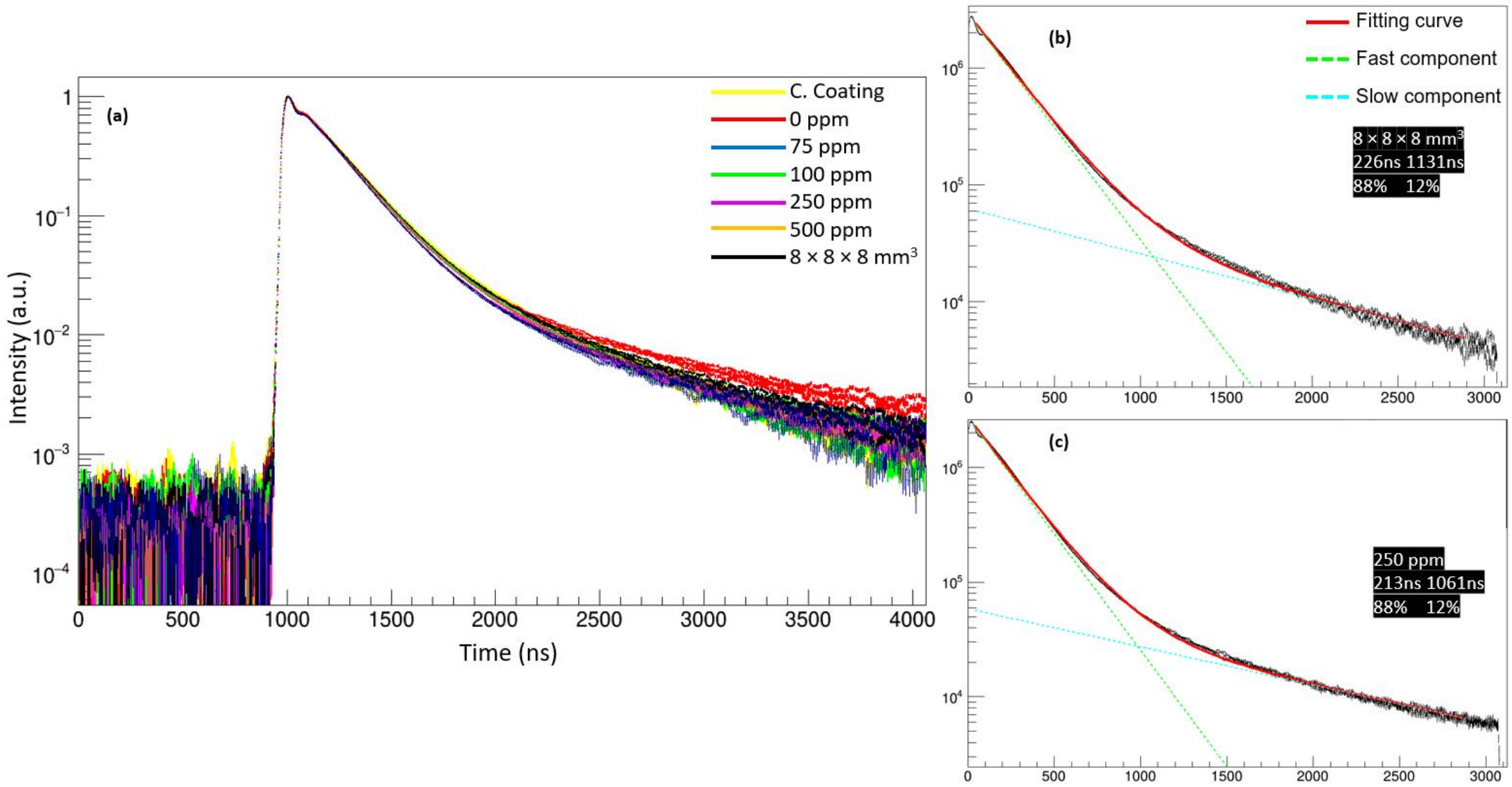

***Fig. 8. (a)*** *Decay times of all small-sized samples compared with the 8 × 8 × 8 mm³ reference crystal;* ***(b)*** *Decay fitting of the reference; and* ***(c)*** *Decay fitting of the 250 ppm sample.*

## 2.4. NaI–quartz interfacial morphology and compositional analysis

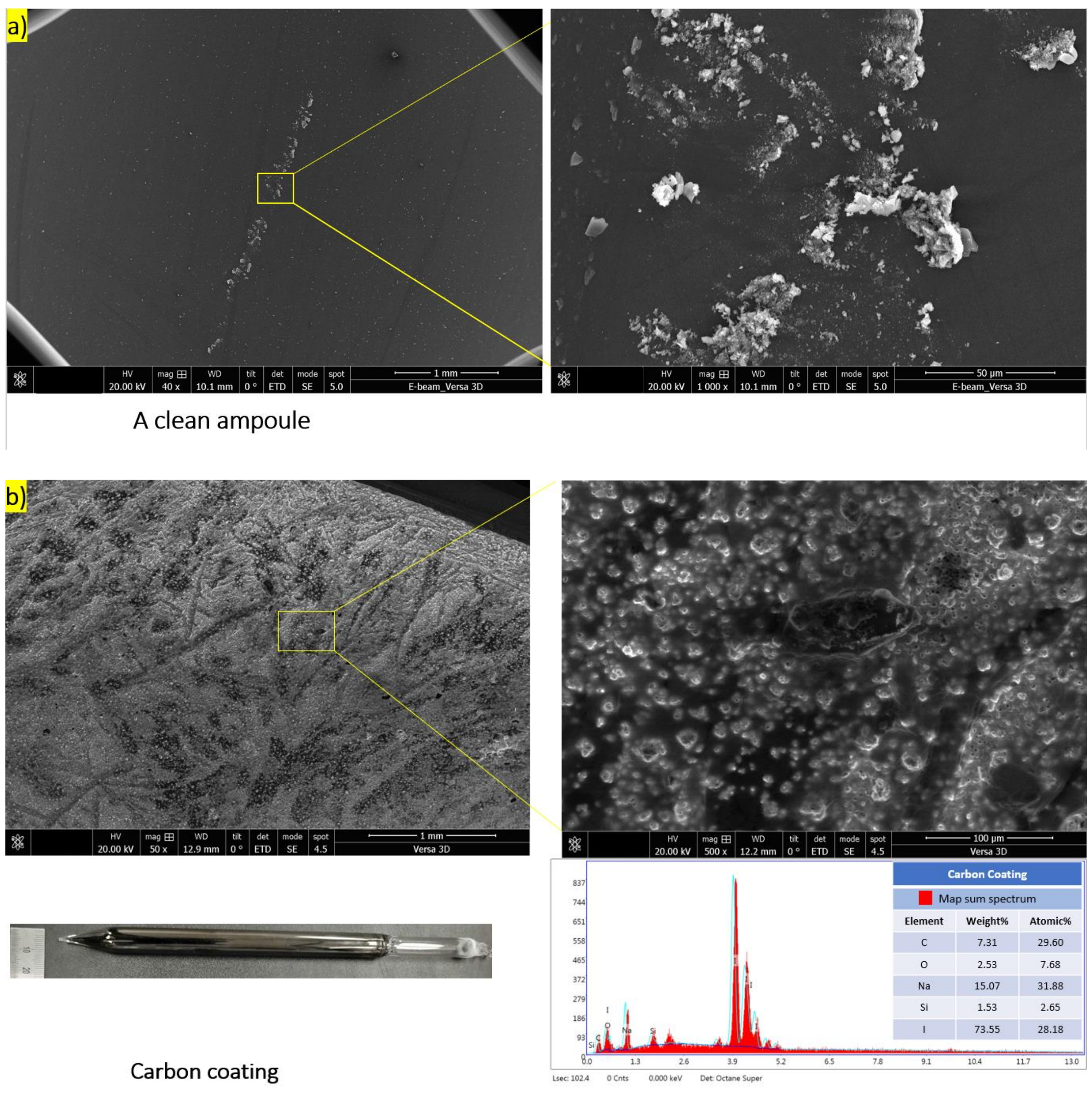

| Carbon Coating | | |
|---|---|---|
| Map sum spectrum | | |
| Element | Weight% | Atomic% |
| C | 7.31 | 29.60 |
| O | 2.53 | 7.68 |
| Na | 15.07 | 31.88 |
| Si | 1.53 | 2.65 |
| I | 73.55 | 28.18 |

c)

| HV | mag | WD | tilt | det | mode | spot | 1 mm |
|---|---|---|---|---|---|---|---|
| 20.00 kV | 50 x | 10.0 mm | 0 ° | ETD | SE | 4.5 | Versa 3D |

| HV | mag | WD | tilt | det | mode | spot | 100 µm |
|---|---|---|---|---|---|---|---|
| 20.00 kV | 500 x | 9.2 mm | 0 ° | ETD | SE | 4.5 | Versa 3D |

0ppm $NH_4I$

**0ppm $NH_4I$** — Map sum spectrum

| Element | Weight% | Atomic% |
|---|---|---|
| C | 3.18 | 15.72 |
| O | 0.77 | 2.84 |
| Na | 17.30 | 44.64 |
| Si | 0.00 | 0.00 |
| I | 78.75 | 36.81 |

Lsec: 102.4 0 Cnts 0.000 keV Det: Octane Super

d)

| HV | mag | WD | tilt | det | mode | spot | 1 mm |
|---|---|---|---|---|---|---|---|
| 20.00 kV | 50 x | 9.2 mm | 0 ° | ETD | SE | 4.5 | Versa 3D |

| HV | mag | WD | tilt | det | mode | spot | 100 µm |
|---|---|---|---|---|---|---|---|
| 20.00 kV | 500 x | 9.7 mm | 0 ° | ETD | SE | 4.5 | Versa 3D |

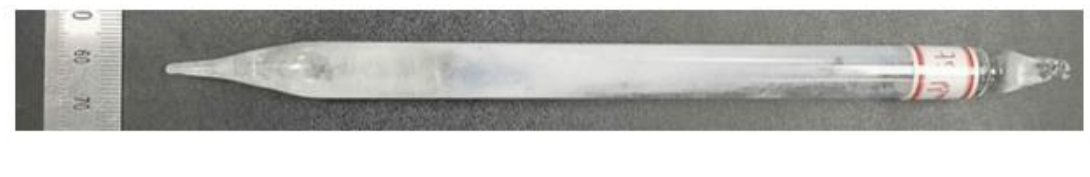

75ppm $NH_4I$

**75ppm $NH_4I$** — Map sum spectrum

| Element | Weight% | Atomic% |
|---|---|---|
| C | 3.79 | 19.27 |
| O | 0.25 | 0.95 |
| Na | 15.39 | 40.89 |
| Si | 0.08 | 0.16 |
| I | 80.49 | 38.73 |

Lsec: 102.4 0 Cnts 0.000 keV Det: Octane Super

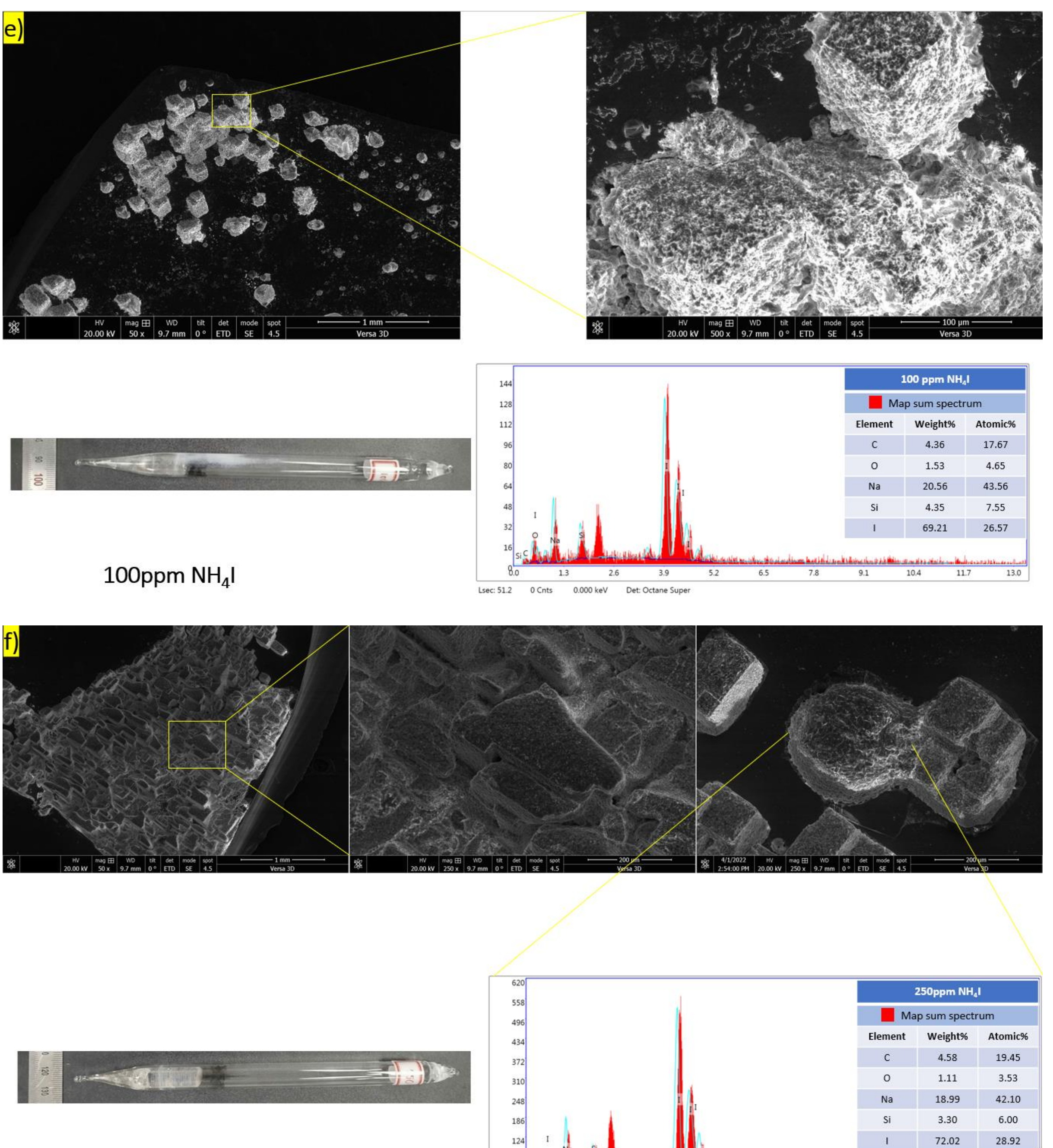

e)
HV 20.00 kV mag 50 x WD 9.7 mm tilt 0 ° det ETD mode SE spot 4.5 1 mm Versa 3D
HV 20.00 kV mag 500 x WD 9.7 mm tilt 0 ° det ETD mode SE spot 4.5 100 μm Versa 3D
100 ppm NH4I
Map sum spectrum
Element | Weight% | Atomic%
C | 4.36 | 17.67
O | 1.53 | 4.65
Na | 20.56 | 43.56
Si | 4.35 | 7.55
I | 69.21 | 26.57
Lsec: 51.2 0 Cnts 0.000 keV Det: Octane Super
100ppm NH4I
f)
HV 20.00 kV mag 50 x WD 9.7 mm tilt 0 ° det ETD mode SE spot 4.5 1 mm Versa 3D
HV 20.00 kV mag 250 x WD 9.7 mm tilt 0 ° det ETD mode SE spot 4.5 200 μm Versa 3D
4/1/2022 2:54:00 PM HV 20.00 kV mag 250 x WD 9.7 mm tilt 0 ° det ETD mode SE spot 4.5 200 μm Versa 3D
250ppm NH4I
Map sum spectrum
Element | Weight% | Atomic%
C | 4.58 | 19.45
O | 1.11 | 3.53
Na | 18.99 | 42.10
Si | 3.30 | 6.00
I | 72.02 | 28.92
Lsec: 92.2 0 Cnts 0.000 keV Det: Octane Super

250ppm $NH_4I$

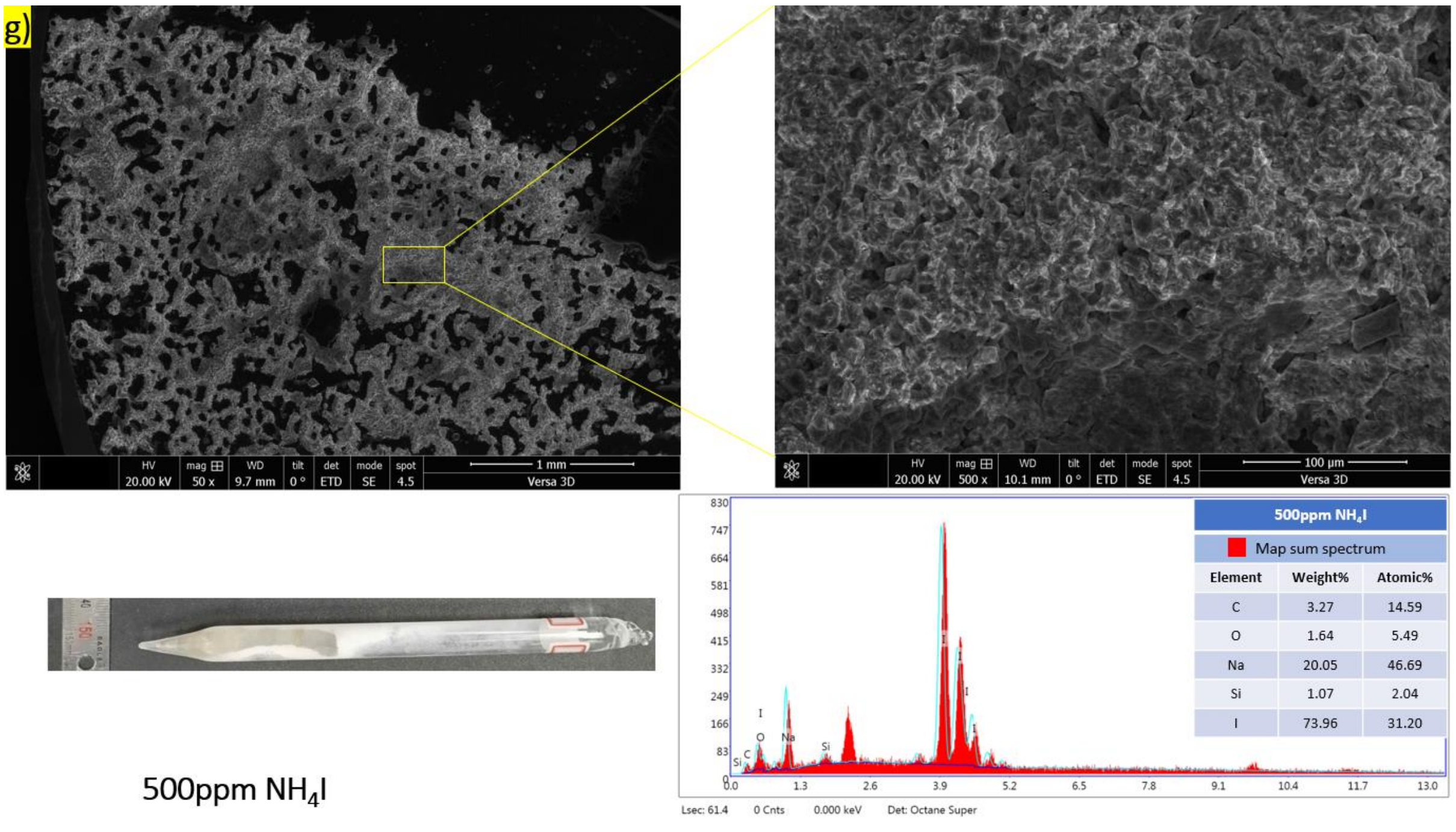

*Fig. 9. SEM images and EDS analysis of the inner surfaces of **(a)** a clean ampoule, **(b)** carbon-coated, **(c)** 0 ppm, **(d)** 75 ppm, **(e)** 100 ppm, **(f)** 250 ppm, and **(g)** 500 ppm samples.*

Since the quartz ampoule is $SiO_2$, contributions from Si and O are expected in the EDS spectra, particularly when the analyzed regions include exposed or partially covered quartz surfaces. As shown in Fig. 9, the spectra are dominated by iodine (I) and sodium (Na), with smaller contributions from oxygen (O) and silicon (Si).

Among the samples with $NH_4I$ concentrations from 0 to 500 ppm, as well as the carbon-coated reference, the EDS results (Fig. 9) show broadly similar elemental compositions, with contributions from I, Na, O, and Si. This similarity is expected because the amount of residual NaOH in the system is relatively small compared to the bulk NaI, making its direct detection by EDS challenging, particularly given the localized nature of the measurements.

A noticeable carbon (C) signal of about 15-20% was observed in all samples, which is most likely attributed to contributions from the carbon tape used for sample mounting, as well as possible surface contamination during handling and SEM measurement.

Despite the limited compositional differences, the SEM images reveal clear variations in surface morphology. In particular, the observed structures differ significantly from the well-defined cubic morphology characteristic of NaI crystals, instead showing irregular features associated with interfacial reaction products.

A plausible interpretation is that residual NaOH at the NaI–quartz interface may react with the $SiO_2$ surface at elevated temperature, forming an interfacial Na–Si–O phase. This interfacial phase may wet the interface and locally disrupt the normal crystallization of NaI, leading to the formation of irregular morphologies rather than well-defined cubic structures. Upon cooling, the interfacial product solidifies and may mechanically anchor the crystal to the quartz wall. The mismatch in thermal expansion between the crystal and this interfacial layer can then contribute to sticking and, in some cases, cracking during cooldown.

To better understand the origin of the interfacial morphology, the relevant chemical reactions are summarized as follows:

$$SiO_2 + 2NaOH \rightarrow Na_2SiO_3 + H_2O$$

As described above, residual NaOH can react with the quartz ($SiO_2$) surface to form sodium–silicate-type products, which are known to contribute to strong adhesion [10, 11].

To suppress this reaction, $NH_4I$ was introduced as an additive with the aim of modifying the chemical environment inside the sealed ampoule. Upon heating, $NH_4I$ may decompose or dissociate to produce gaseous species, potentially including $NH_3$ and HI:

$$NH_4I \rightarrow NH_3\ (g) + HI\ (g)$$

Once HI is formed, it can react with NaOH, leading to the formation of NaI and reducing the availability of NaOH for reaction with quartz:

$$HI + NaOH \rightarrow NaI + H_2O$$

To further investigate the formation mechanism of the NaI–quartz interfacial morphology, dedicated ampoule experiments, as described in Section 1.2, were performed (Fig. 3). The results are presented below.

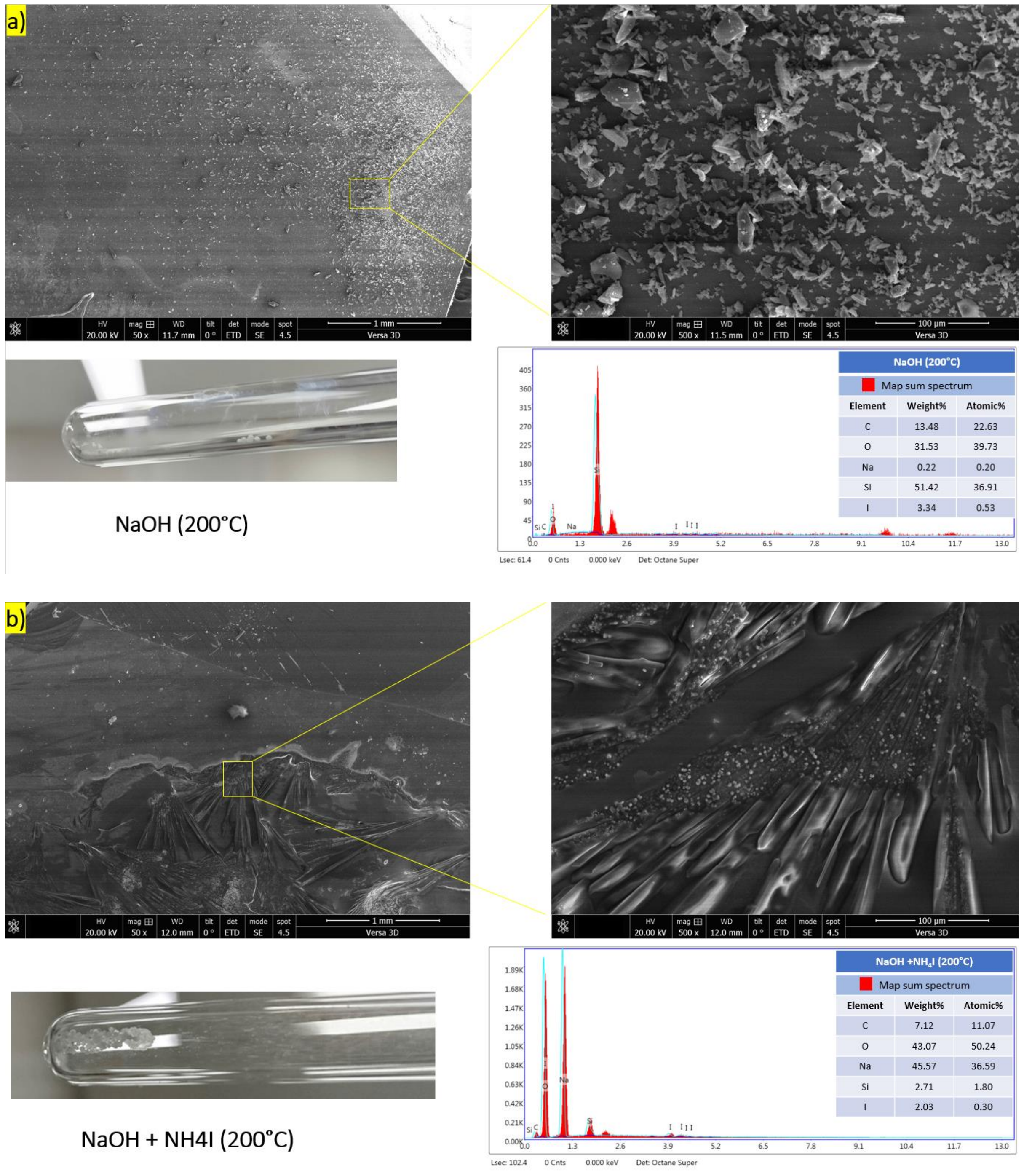

a)
HV 20.00 kV mag 50 x WD 11.7 mm tilt 0 ° det ETD mode SE spot 4.5 1 mm Versa 3D
HV 20.00 kV mag 500 x WD 11.5 mm tilt 0 ° det ETD mode SE spot 4.5 100 µm Versa 3D
NaOH (200°C)
NaOH (200°C)
Map sum spectrum
Element Weight% Atomic%
C 13.48 22.63
O 31.53 39.73
Na 0.22 0.20
Si 51.42 36.91
I 3.34 0.53
Lsec: 61.4 0 Cnts 0.000 keV Det: Octane Super
b)
HV 20.00 kV mag 50 x WD 12.0 mm tilt 0 ° det ETD mode SE spot 4.5 1 mm Versa 3D
HV 20.00 kV mag 500 x WD 12.0 mm tilt 0 ° det ETD mode SE spot 4.5 100 µm Versa 3D
NaOH +NH4I (200°C)
Map sum spectrum
Element Weight% Atomic%
C 7.12 11.07
O 43.07 50.24
Na 45.57 36.59
Si 2.71 1.80
I 2.03 0.30
Lsec: 102.4 0 Cnts 0.000 keV Det: Octane Super
NaOH + NH4I (200°C)

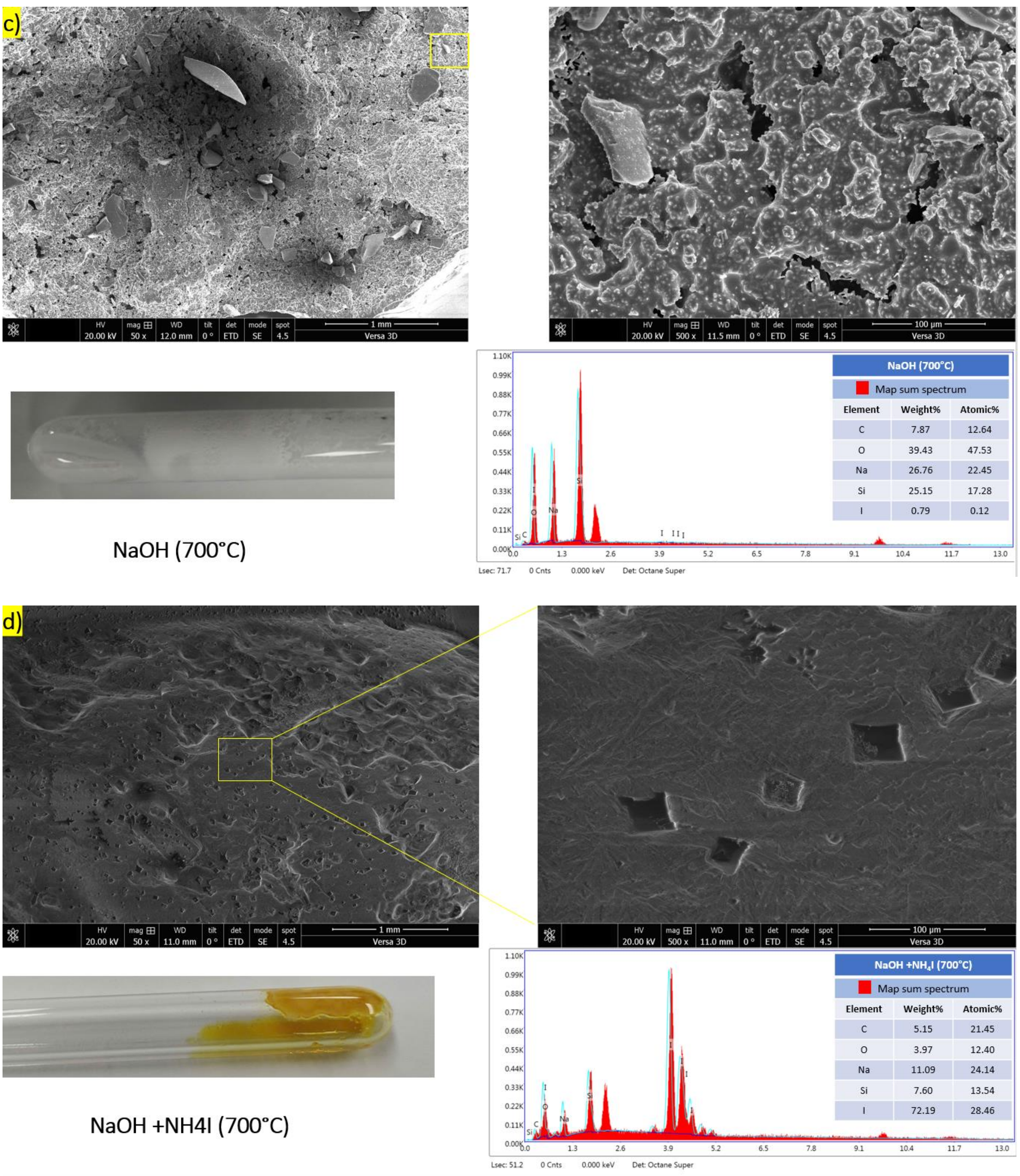

***Fig. 10.*** *SEM images and EDS analysis of the inner surface of the* ***(a)*** *NaOH ampoule at 200 °C,* ***(b)*** *NaOH +$NH_4I$ at 200 °C,* ***(c)*** *NaOH ampoule at 700°C, and* ***(d)*** *NaOH +$NH_4I$ at 700°C.*

Fig. 10 shows the results of the NaOH and NaOH + $NH_4I$ ampoule tests at 200 °C and 700 °C. In the NaOH-only sample at 200 °C (Fig. 10 (a)), no significant change was observed, as the NaOH beads remained free to move within the ampoule. Since the melting point of NaOH is approximately 320 °C, the material remains in the solid phase at 200 °C, and reaction with the quartz surface is expected to be limited. Consistent with this, EDS analysis shows primarily Si and O signals primarily, corresponding to the quartz substrate.

In the NaOH + $NH_4I$ sample at 200 °C (Fig. 10 (b)), noticeable changes in bead morphology were observed. The NaOH particles exhibited deformation and partial adhesion to the quartz surface. EDS analysis of these regions revealed the presence of I in addition to Na, O, and Si, indicating interaction between $NH_4I$-derived species and NaOH. Under these conditions, partial sublimation or decomposition of $NH_4I$ may occur, producing reactive gaseous species that interact with the surface of solid NaOH. However, since NaOH remains in the solid phase, the overall reaction is likely limited. The I signal also shows a consistently small portion.

In contrast, the NaOH-only sample at 700 °C (Fig. 10(c)) exhibited strong adhesion to the quartz surface, forming a solid, rigid interfacial product after cooling. EDS analysis shows dominant contributions from Na, Si, and O, consistent with the formation of sodium–silicate products from the reaction between molten NaOH and $SiO_2$.

For the NaOH + $NH_4I$ sample at 700 °C (Fig. 10 (d)), the interfacial product was observed to be less rigid and less strongly adhered to the quartz surface. This qualitative observation is supported by EDS results, which show a dominant I signal, while the relative contributions of Na, Si, and O are reduced. These results suggest that the presence of $NH_4I$ is associated with changes in the interfacial reaction pathway, leading to iodine-containing interfacial products and reducing the formation of the strongly adhesive Na-Si-O phase.

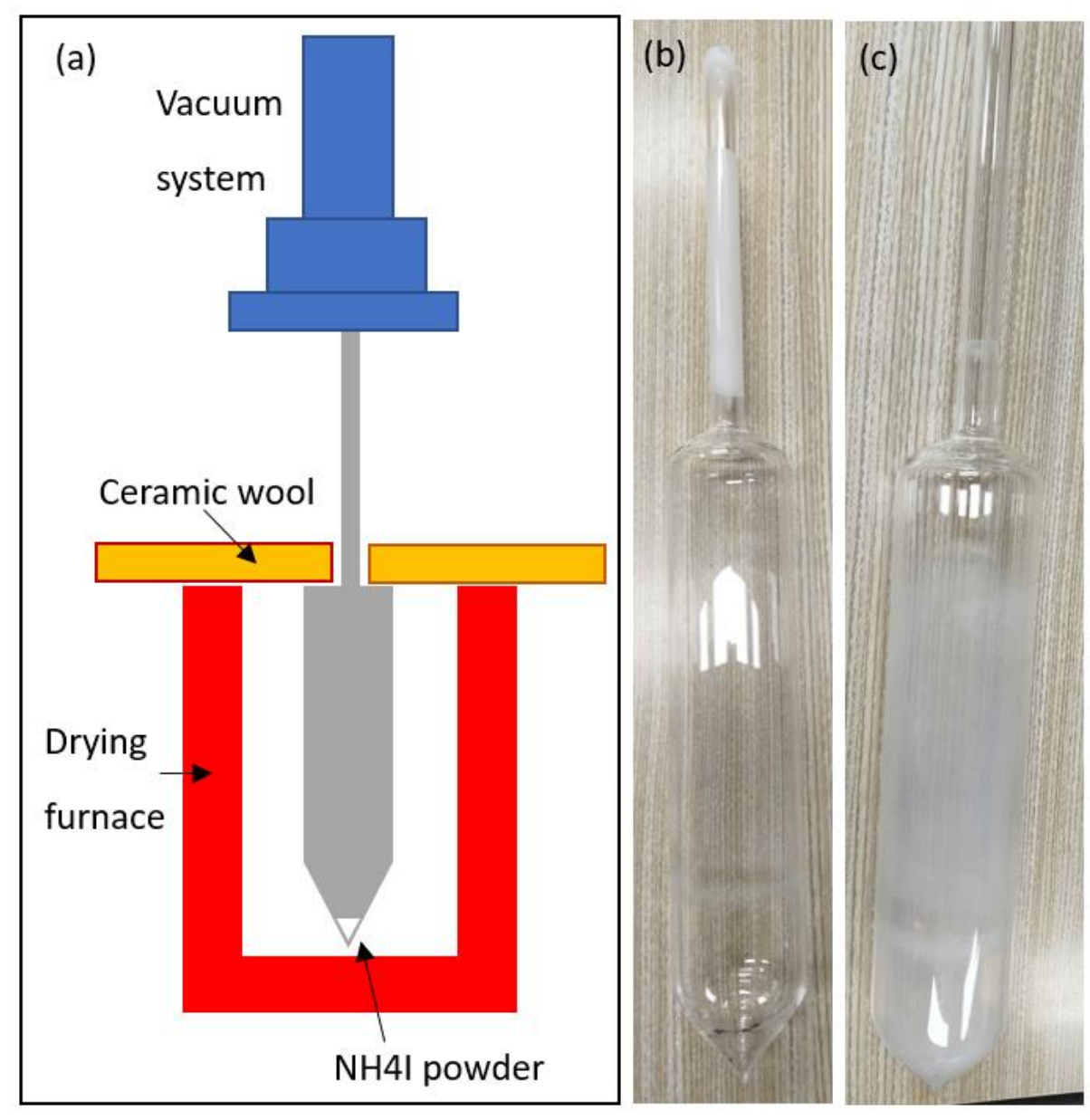

*Fig. 11.* ***(a)*** *Drying and vacuum system;* ***(b)*** *$NH_4I$ condensed at the top of the ampoule after sublimating at 140 °C during drying;* ***(c)*** *after sealing and re-heating upside down at 140 °C, $NH_4I$ sublimed again and condensed near the ampoule tip.*

Additional evidence for the role of $NH_4I$ was obtained from controlled tests performed during the drying process. At 140 °C under a high vacuum (~$10^{-6}$ Torr), partial sublimation of $NH_4I$ was observed, as indicated by the condensation of $NH_4I$ at the cooler regions of the ampoule (Fig. 11). This behavior is attributed to the reduced pressure, which enhances the volatility of $NH_4I$ and enables sublimation at temperatures below its nominal boiling point. As a result, a portion of $NH_4I$ is lost during drying and sealing, while a fraction remains within the sealed system. Although this test was performed without NaI, it provides qualitative evidence of $NH_4I$ volatility under the same thermal and vacuum conditions used during ampoule preparation. Even though the thermal behavior of $NH_4I$ was not directly measured in this study, ammonium salts commonly undergo proton-transfer-driven thermal dissociation, and ammonium halides can exhibit sublimation/dissociation behavior upon heating [25-26]. Together with our direct observation of $NH_4I$ condensation in cooler regions of the ampoule, this supports the plausibility of $NH_4I$ volatilization and redistribution during the drying and heating steps.

In an actual NaI(Tl) crystal growth, NaOH can already be present in the system due to prior hydrolysis of NaI during handling or storage, and upon further heating, when the temperature exceeds its melting point (~320 °C), molten NaOH can react with the quartz surface, forming Na–Si–O-type interfacial products. In the absence of sufficient $NH_4I$, this interfacial phase is expected to interact with the NaI melt. During solidification of NaI (~660 °C), the interfacial product likely remains in a molten or viscous state, allowing it to wet the crystal surface. Upon cooling, this interfacial layer solidifies, becomes strongly adherent, and, due to thermal expansion mismatch with NaI, can lead to mechanical stress or cracking.

In contrast, in ampoules containing $NH_4I$, thermal decomposition produces $NH_3$ and HI, which can diffuse throughout the system and are expected to react with NaOH. As demonstrated in the 700 °C reaction experiments, the presence of $NH_4I$ significantly alters the interfacial chemistry, leading to the presence of iodine-containing species and a reduced contribution from Na–Si–O interfacial products. This modification of the interfacial reaction pathway is consistent with the observed suppression of sticking in the $NH_4I$-treated samples. Although the presence of HI was not directly confirmed in this study, the observed changes in interfacial morphology and composition are consistent with this mechanism.

### 2.5. Scale-up to large-size and performance

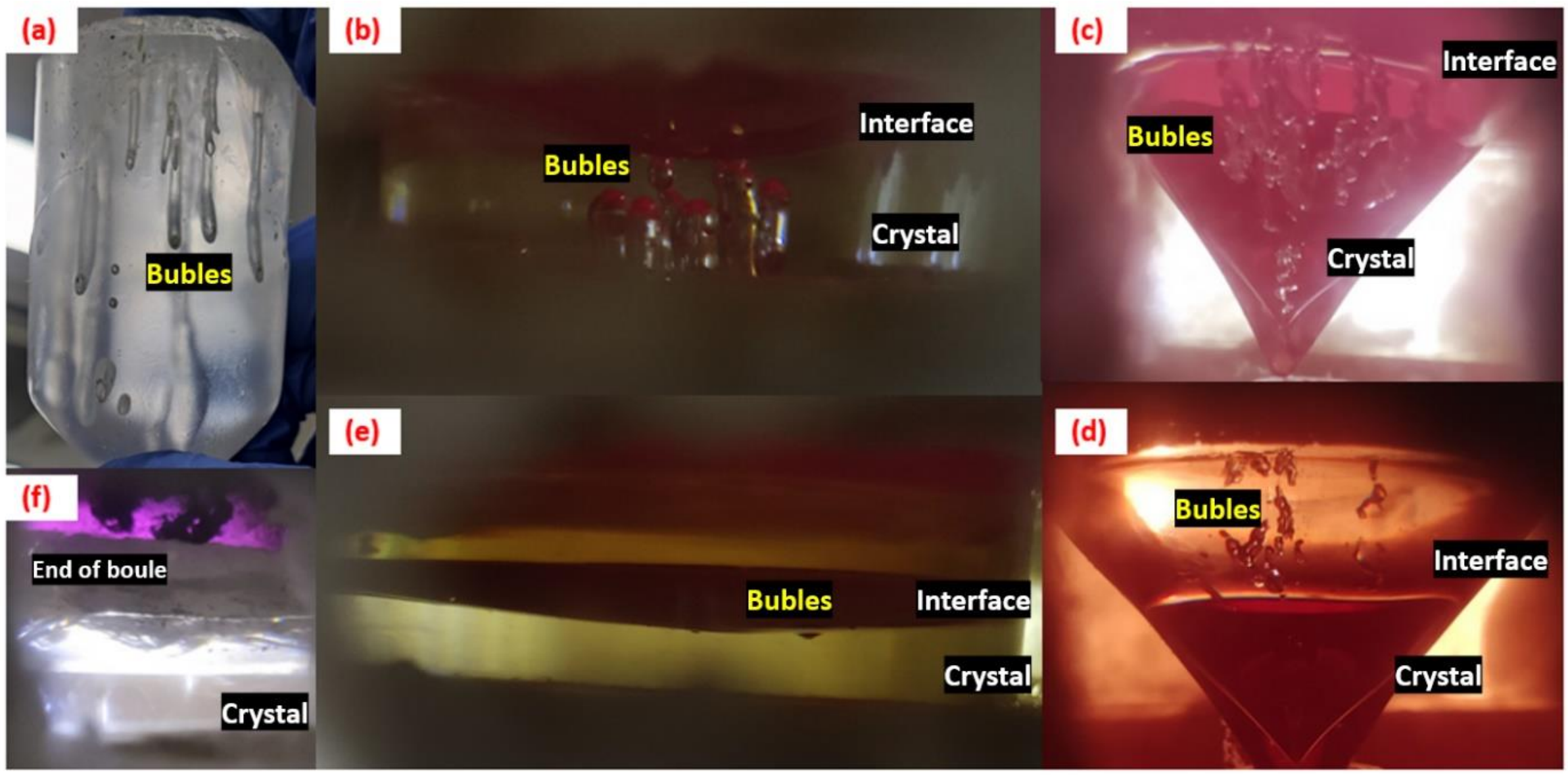

***Fig. 12.*** ***(a)*** *Room-temperature 2-inch crystals grown at 1.2 mm/h, containing internal bubbles; observed solid–liquid interface through a window at* ***(b, c)*** *1.0 mm/h,* ***(d)*** *0.8 mm/h, and* ***(e)*** *0.6 mm/h showing a small trapped bubble.* ***(f)*** *Violet coloration attributed to iodine-containing vapor species was observed during cooling at 300–400 °C.*

With adhesion suppression and scintillation performance confirmed at small scale, the optimized $NH_4I$ condition was applied to the growth of large-diameter NaI(Tl) crystals. Based on the trends, an intermediate $NH_4I$ concentration of 200 ppm was selected. All other growth parameters were kept constant, except that the drying temperature was lowered from 140 °C to 100 °C to reduce $NH_4I$ loss due to sublimation under high vacuum.

During the scaling up, heat transfer within the melt became more complex, and bubble formation became increasingly difficult to suppress. In this work, bubble-related defects were observed in both small- and large-size growths; however, their presence increased with increasing crystal diameter. This behavior is relevant because larger NaI charges require more $NH_4I$, which is expected to decompose during heating and may introduce additional gaseous species into the melt. As a result, while a growth rate of approximately 1.0–1.2 mm/h was sufficient to allow bubble release in small-size crystals, significantly lower growth rates were required for bubble escape from the solid–liquid interface. Similar bubble behavior during NaI growth has been reported in previous studies [6,9]. Berthold et al. [9] demonstrated that bubble trapping can be effectively suppressed by reducing the growth rate. Following this approach, the growth rate in this work was reduced from 1.2 mm/h, used for small-size samples, to 0.8–0.6 mm/h for large-diameter crystals (2″ and 3″). As shown in Fig. 12, decreasing the growth rate significantly reduced bubble trapping at the solid–liquid interface. At a growth rate of 0.6 mm/h, both 2-inch and 3-inch NaI(Tl) crystals could be grown bubble-free. The successfully grown 3 x 3-inch NaI(Tl) crystal is shown in Fig. 13.

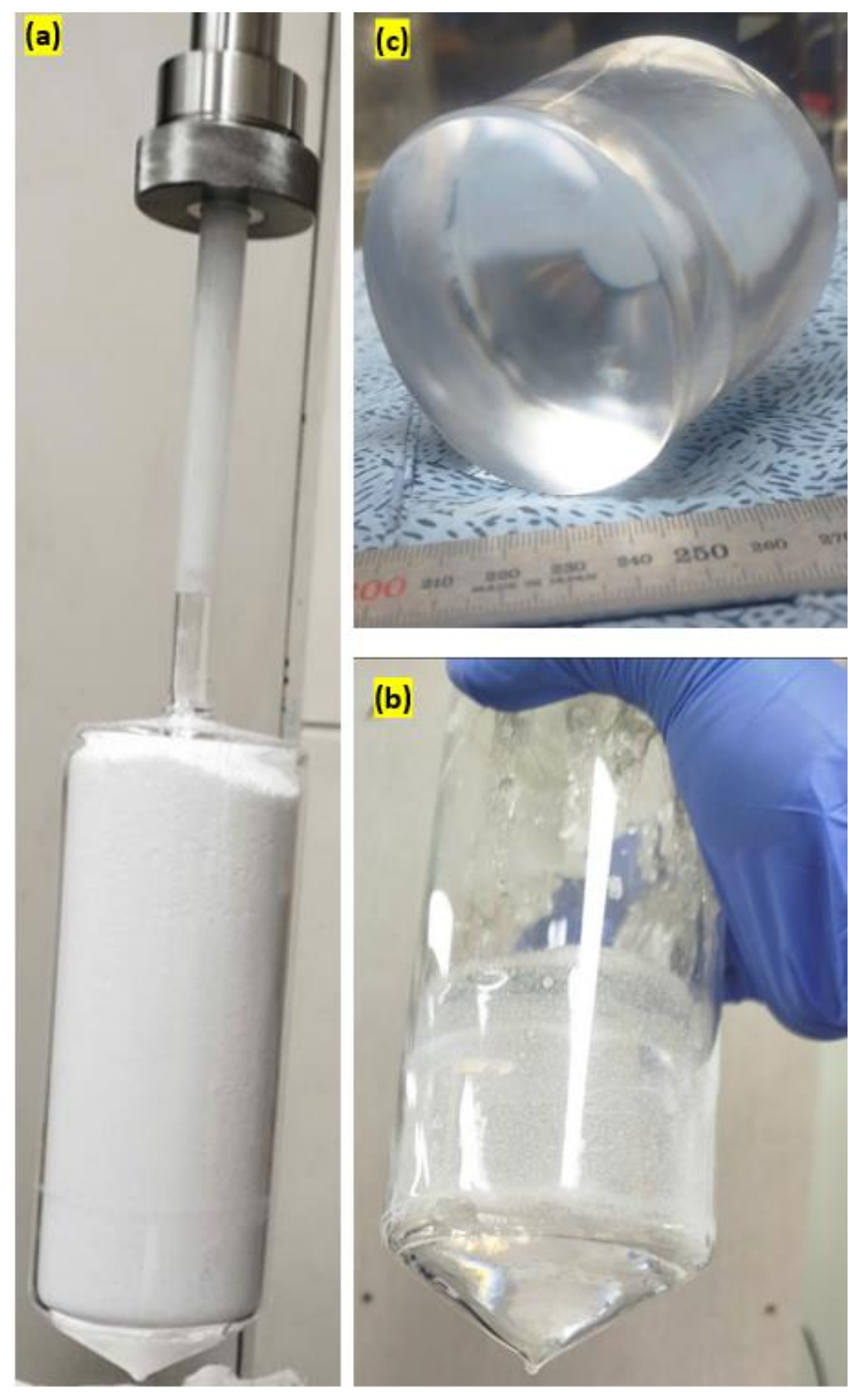

***Fig. 13.*** *Φ3-inch x 3-inch crack-free, bubble-free NaI(Tl) crystal grown using the $NH_4I$ treatment.* ***(a)*** *Ampoule under vacuum before sealing;* ***(b)*** *right after growth;* ***(c)*** *Φ3-inch x 3-inch, 0.9 kg crystal after simple polishing.*

Figure 14 (a) compares the pulse height spectrum of the $NH_4I$-treated 3-inch NaI(Tl) crystal with those of commercial 2-inch and 3-inch NaI(Tl) crystals from EPIC and Alpha Spectra, measured under $^{137}Cs$ (662 keV) γ-ray excitation. The Gaussian function-fitted photopeak energy resolution of the $NH_4I$-treated crystal is shown in Fig. 14(b). The $NH_4I$-treated 3-inch crystal exhibits a noticeably higher light output than both commercial reference detectors, indicating good bulk optical quality and efficient scintillation light collection. The detailed scintillation parameters are summarized in Table 2.

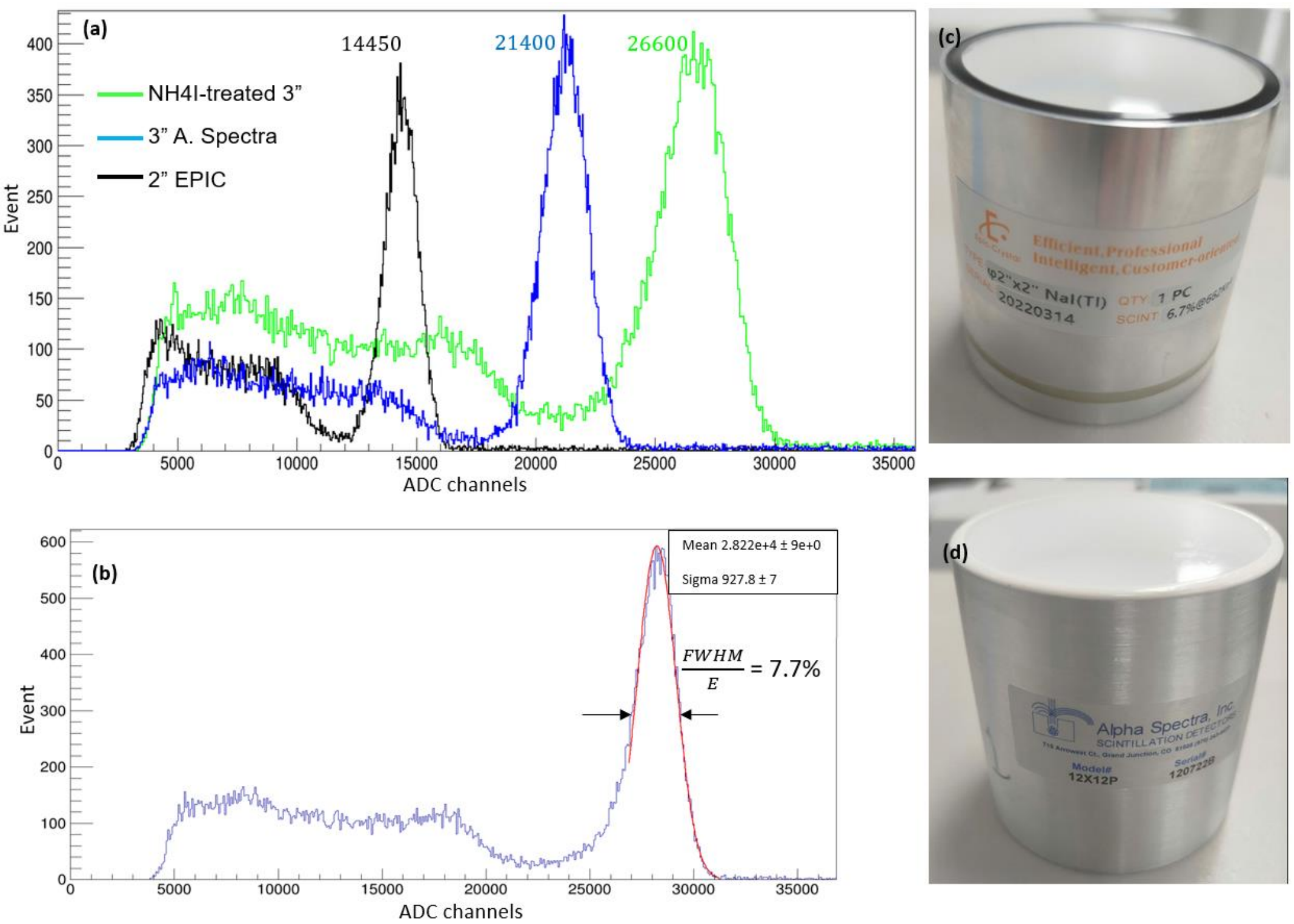

***Fig. 14.*** ***(a)*** *Pulse height spectra at 662 keV under* $^{137}$*Cs γ-ray excitation for the* $NH_4I$*-treated 3-inch crystal compared with commercial reference crystals;* ***(b)*** *Energy resolution of the* $NH_4I$*-treated 3-inch crystal;* ***(c)*** *2-inch EPIC and* ***(d)*** *3-inch A. Spectra crystal.*

Although the $NH_4I$-treated 3-inch crystal shows slightly poorer energy resolution than commercial references, this difference is primarily attributed to the non-optimized geometry of the tested sample, which was not fully machined into a standard symmetric shape, rather than intrinsic scintillation degradation. With the $NH_4I$ treatment, the recovery rate of usable crystal material was nearly 100 %, as the ingot could be extracted intact from the quartz ampoule. This high recovery ratio demonstrates the robustness and scalability of the $NH_4I$-based approach for large-size NaI(Tl) crystal growth.

***Table 2****: The measured energy resolution (FWHM/E) and light yield at 662 keV of 2 and 3-inch NaI(Tl) commercial crystals compared with the 3-inch NH4I sample.*

| **NaI(Tl) sample** | **Light yield** (ph/MeV) | **Energy Resolution** (FWHM/E) (%) |
|---|---|---|
| 2-inch EPIC | 32,400 ± 3,240 [15] | 6.7 ± 0.1 |
| 3-inch A. Spectra | 48,000 ± 4,800 [15] | 6.5 ± 0.1 |
| $NH_4I$-treated 3-inch | 59,000 ± 5,900 (this work) | 7.7 ± 0.1 |

# III. Conclusion

Using a small amount of $NH_4I$ as an additive, we successfully grew crack-free and stick-free bulk NaI(Tl) crystals by the vertical Bridgman method in quartz ampoules. The addition of $NH_4I$ effectively suppressed adhesion between the crystal and the quartz wall without introducing foreign coating materials.

SEM/EDS analysis and dedicated NaOH-based ampoule experiments indicate that adhesion originates from interfacial reactions between residual NaOH and quartz, leading to the formation of Na–Si–O-type products. The results also suggest that $NH_4I$ modifies this interfacial reaction pathway, promoting the formation of iodine-rich products and significantly reducing the formation of strongly adhesive phases.

This approach requires no additional processing steps and can be implemented simply by mixing $NH_4I$ with the starting materials, making it practical for laboratory-scale crystal growth. Adhesion was completely eliminated under optimized conditions, and scintillation performance remained within the expected range for laboratory-grown crystals. The resulting crystal showed a light yield of 59,000 ph/MeV, which is higher than other commercial products used as references. Nearly the entire grown boule was recovered, representing a significant advantage for applications using high-cost ultra-pure NaI powder.

For the present system, the optimal $NH_4I$ concentration was approximately 200 ppm, although this value may vary depending on material purity, size of the crystal, ampoule condition, and growth parameters. A minor side effect of the treatment is increased bubble formation, which can be mitigated by reducing the growth rate to 0.8–0.6 mm/h.

Overall, this additive-based strategy provides a clean and effective alternative to conventional coating methods and may potentially be extended to other alkali-halide systems with similar quartz adhesion mechanisms. It should be noted that the results presented in this work are based on ultra-high-purity NaI starting materials. Therefore, further validation using commercially available NaI powders is required to assess the general applicability of this approach.

**Acknowledgments** This work was supported by the Institute for Basic Science (IBS) and the National Research Foundation of Korea (NRF), Grant funded by the Korean government (MSIP) and Ministry of Science and Technology (MEST), (RS-2025-25460489, RS-2024-00348317 and RS-2018-NR031074, RS-2024-00447038)

## References

[1] G.F. Knoll, Radiation Detection and Measurement, fourth ed., Wiley, Hoboken, 2010.

[2] W.J. Van Sciver, Fluorescence and reflection spectra of NaI single crystals, Phys. Rev. 120 (1960) 1193–1205. https://doi.org/10.1103/PhysRev.120.1193

[3] R. Hawrami, E. Ariesanti, A. Farsoni, D. Szydel, H. Sabet, Growth and evaluation of improved CsI:Tl and NaI:Tl scintillators, Crystals 12 (2022) 1517. https://doi.org/10.3390/cryst12111517

[4] B.J. Park et al., Development of ultra-pure NaI(Tl) detectors for the COSINE-200 experiment, *Eur. Phys. J. C* 80 (2020) 814. https://doi.org/10.1140/epjc/s10052-020-8386-8

[5] B. Suerfu et al., Growth of ultra-high purity NaI(Tl) crystals for dark matter searches, *Phys. Rev. Research* 2 (2020) 013223. https://doi.org/10.1103/PhysRevResearch.2.013223

[6] B. Suerfu, *Developing Ultra-Low Background Sodium-Iodide Crystal Detector for Dark Matter Searches*, Ph.D. thesis, Princeton University, 2018.

[7] K. Shin et al., A facility for mass production of ultra-pure NaI powder for the COSINE-200 experiment, *J. Instrum.* 15 (2020) C07031. https://doi.org/10.1088/1748-0221/15/07/C07031

[8] S.G. Singh, R. Kumar, H.C. Verma, Growth of Tl-doped CsI and NaI single crystals in a modified Bridgman furnace, *DAE Symp. Nucl. Phys.* 59 (2000) 914–915.

[9] T. Berthold, M. Hofmann, W. Schötzig, Growth of single crystalline NaI plates, *J. Cryst. Growth* 217 (2000) 441–448. https://doi.org/10.1016/S0022-0248(00)00487-5

[10] V.S. Balitsky et al., Kinetics of dissolution and state of silica in $Na_2CO_3$ and NaOH hydrothermal solutions, *J. Cryst. Growth* 237–239 (2002) 828–832. https://doi.org/10.1016/S0022-0248(01)02045-0

[11] A.M. Ali et al., Molecular simulation and microtextural characterization of quartz dissolution in sodium hydroxide, *J. Petrol. Explor. Prod. Technol.* 10 (2020) 2669–2684. https://doi.org/10.1007/s13202-020-00940-2

[12] J.H. Mark, D. Miller, R. Wang, Carbon coating of fused silica ampoules, *J. Cryst. Growth* 290 (2006) 597–601. https://doi.org/10.1016/j.jcrysgro.2006.01.014

[13] B.R. Johnson et al., FY06 Annual Report: Amorphous Semiconductors for Gamma Radiation Detection (ASGRAD), 2006. https://doi.org/10.2172/1047430

[14] Q.V. Phan et al., Development of novel crystal scintillators for lunar surface science, *Radiat. Phys. Chem.* 201 (2022) 110425. https://doi.org/10.1016/j.radphyschem.2022.110425

[15] L.T. Truc et al., Scintillation properties of cerium-doped $Tl_2LaCl_5$ crystals, *Radiat. Phys. Chem.* 237 (2025) 113059. https://doi.org/10.1016/j.radphyschem.2025.113059

[16] N.T. Luan et al., Absolute light yield measurement of NaI:Tl crystals for dark matter search, arXiv:2404.18551 (2024). https://arxiv.org/abs/2404.18551

[17] R.C. Merrill et al., Gelation of sodium silicate: effects of sulfuric acid, hydrochloric acid, ammonium sulfate, sodium aluminate, *J. Phys. Colloid Chem.* 54 (1950) 806–812. https://doi.org/10.1021/j150480a009

[18] M. Matinfar et al., A review of sodium silicate solutions: structure, gelation and syneresis, *Adv. Colloid Interface Sci.* 322 (2023) 103036. https://doi.org/10.1016/j.cis.2023.103036

[19] V. Kumar, Z. Luo, A review on X-ray excited emission decay dynamics in inorganic scintillators, *Photonics* 8 (2021) 71. https://doi.org/10.3390/photonics8030071

[20] F.S. Eby, W.K. Jentschke, Fluorescent response of NaI(Tl) to nuclear radiations, *Phys. Rev.* 96 (1954) 911–920. https://doi.org/10.1103/PhysRev.96.911

[21] J. Wang et al., Segregation effects of Tl and Li ions on scintillation properties of NaI:Tl,Li crystals, *J. Cryst. Growth* 612 (2023) 127199. https://doi.org/10.1016/j.jcrysgro.2023.127199

[22] Y. Zhu et al., Production of ultra-low radioactivity NaI(Tl) crystals for dark matter search, https://arxiv.org/pdf/1909.11692 (2019)

[23] Robert S. Feigelson, Crystal growth History: Theory and melt growth processes, *J. of Cryst. Growth* 594 (2022) 126800, ISSN 0022-0248, https://doi.org/10.1016/j.jcrysgro.2022.126800

[24] Shin K, Choe J, Gileva O, Iltis A, Kim Y, Kim Y, Lee C, Lee E, Lee H and Lee MH (2023) Mass production of ultra-pure NaI powder for COSINE-200. *Front. Phys*. 11:1142849. https://doi.org/10.3389/fphy.2023.1142849

[25] Andrew K. Galwey, Michael E. Brown, Chapter 15 Decomposition of ammonium salts, *Studies in Physical and Theoretical Chemistry*, Elsevier Vol. 86 (1999), Pages 415-440, ISSN 0167-6881, ISBN 9780444824370, https://doi.org/10.1016/S0167-6881(99)80017-2

[26] M. Olszak-Humienik, On the thermal stability of some ammonium salts, *Thermochimica Acta*,Vol. 378, Issues 1–2 (2001) Pages 107-112, ISSN 0040-6031, https://doi.org/10.1016/S0040-6031(01)00585-8

[27] A. Kozlov et al., Detectors for direct Dark Matter search at KamLAND, Nuclear Instruments and Methods in Physics Research Section A: Accelerators, Spectrometers, Detectors and Associated Equipment, Vol. 958 (2020), ISSN 0168-9002, https://doi.org/10.1016/j.nima.2019.05.080

[28] Smausz et al., Determination of UV–visible–NIR absorption coefficient of graphite bulk using direct and indirect methods. Appl. Phys. A 123, 633 (2017), https://doi.org/10.1007/s00339-017-1249-y